\documentclass[12pt,manuscript]{aastex}

\usepackage{natbib} 

\begin{document} 

\title{Relative velocities among accreting planetesimals in binary systems:\\
the circumprimary case}

\author{P. Th\'ebault}
\affil{ Stockholm Observatory, Albanova Universitetcentrum,
SE-10691 Stockholm, Sweden\\
and}
\affil{ 
Observatoire de Paris, Section de Meudon, F-92195 Meudon Principal Cedex,
France} 
\email{email: philippe.thebault@obspm.fr}
\author{F.Marzari}
\affil{Dipartimento di Fisica, Universita di Padova, Via Marzolo 8, I-35131 
Padova, Italy}
\and
\author{H. Scholl}
\affil{
Observatoire de la C\^ote d'Azur, Dept. Cassiop\'ee, B.P. 4229, 
F-06304 Nice, France}

Manuscript Pages: 31

Tables: 4 

Figures: 10

\clearpage

\textbf{Proposed Running Head:}planetesimal accretion in binaries

\vspace{2cm}

\textbf{Editorial correspondence to:}

Philippe Th\'ebault

Stockholm Observatory

Albanova Universitetcentrum

SE-10691 Stockholm

Sweden

Phone: 46 8 5537 85 59

E-mail: philippe.thebault@obspm.fr

\clearpage

{\bf{ABSTRACT}}

We investigate classical planetesimal accretion in a binary star system
of separation $a_b\leq$50\,AU by numerical simulations, with particular
focus on the region at a distance of 1 AU from the primary.
The planetesimals orbit the primary, are perturbed by the companion and are
in addition subjected to a gas drag force.
We concentrate on the problem of relative velocities 
$\Delta v$ among planetesimals of different sizes.
For various stellar mass ratios and binary orbital parameters we
determine regions where $\Delta v$ exceed planetesimal escape velocities $v_{esc}$
(thus preventing runaway accretion) or even the threshold velocity $v_{ero}$
for which erosion dominates accretion. 
Gaseous friction has two crucial effects on the velocity distribution:
it damps secular perturbations by forcing periastron
alignment of orbits, but at the same time the size--dependence of
this orbital alignment induces a significant $\Delta v$ increase between
bodies of different sizes. 
This differential phasing effect proves very efficient and almost always
increases $\Delta v$ to values preventing runaway accretion, except
in a narrow $e_b \simeq 0$ domain.
The erosion threshold $\Delta v > v_{ero}$ is reached
in a wide ($a_b,e_b$) space for small $<10\,$km planetesimals,
but in a much more limited region for bigger $\simeq50\,$km objects.
In the intermediate $v_{esc}<\Delta v < v_{ero}$ domain, 
a possible growth mode would be the type II runaway growth identified
by \citet{kort01}.

{\bf keywords} planetary formation -- 
        planetary dynamics -- 
        accretion 
                                           
\clearpage

 
%

\section{INTRODUCTION} 
 
The problem of planetary formation in binary systems is a crucial one,
as a majority of solar type stars are believed to reside in binary or
multiple star systems. 
New light on this issue has been shed by the recent discoveries of several
extrasolar planets around stars in binary systems \citep{egg03}.
In the classical planetary formation scenario,
one crucial stage is the
mutual accretion of kilometre-sized planetesimals leading, through runaway
and possibly oligarchic growth, to the formation of lunar--to--Mars sized embryos
on relatively short timescales of $5\times10^{3}$ to $10^{5}\,$ years.
\citep[e.g.][]{gre78,wet89,bar93,lis93,kok98,kok00,raf03,raf04}.
We focus here on the specific problem of planetesimal accretion around
the circumprimary star under the perturbing influence of the companion.
The crucial parameter for this stage is the encounter
velocity between impacting objects, which has to be lower than the
bodies escape velocity (corrected by an energy dissipation factor) in
order to allow mutual accretion. 
In a gravitationally unperturbed disk, this condition is met for a large
fraction of mutual impacts \citep[e.g][]{saf69,gre78},
but external gravitational perturbations might lead to relative
velocity increase and thus inhibit accretion.
Such a risk obviously exists in a binary system, where
the companion star affects the inner disk through
secular perturbations leading to substantial eccentricity oscillations
\citep[e.g.][hereafter TH04]{mascho00,the04}.
However, these secular effects come with a strong forced
orbital phasing between neighboring objects. This tends to keep relative
velocities at a lower level than expected from using
the usual approximation $\Delta v \propto e\,v_{kep}$,
which is anyway often misleading \citep[see for instance the discussion for
the specific case of the Kuiper Belt in][]{the03}.
Nevertheless, the secular oscillations get narrower with time
so that at some point neighboring orbits eventually cross,
leading to very high encounter velocities.
This orbital crossing criterion has been explored analytically
by \citet{hep78}, who used a simplified estimate for the apsidal line
regression timescale. \citet{whit98} carried out a more detailed study
by numerically integrating the orbits of the stars of the binary system and of 
two massless planetesimals over a typical runaway growth timescale of
$2\times10^{4}$\,years. This study claims that, within the timescale
of their numerical simulations, the main variations of 
the planetesimal orbital elements  
are due to short term effects of the binary gravity field
that would be averaged out in a secular theory.
However, it is not clear if their adopted methodology, i.e. extrapolating
orbital crossing criteria from orbital parameters recorded only
at binary periastron, allows to correctly estimate these short term
intra-orbit effects. In addition, as the authors point out themselves, 
the choice of initial parameters and assumptions were made in order to
get the most conservative criteria for orbital crossing, thus 
necessarily overestimating the short term effect's efficiency.
Furthermore, over $2\times10^{4}$\,years the secular perturbations 
have sufficient time to build up a dephasing $d\varpi$ in the longitude
of periastron of two planetesimals. This $d\varpi$ depends 
on the initial difference in semimajor axis between the orbits of the 
two bodies. Unfortunately, it is difficult for their model to separate
the contribution to the impact velocity of the short term effects from 
that coming from the dephasing of the secular perturbations. 
Finally, the potentially crucial effect of gas drag on relative velocity
evolution was left out of this purely gravitational study.

We shall here adopt a different approach, based on the numerical
scheme developed in earlier studies \citep[e.g.][]{mascho00,the02,the04}
to study the encounter velocity evolution, within a test planetesimal population,
under the coupled effect of dynamical perturbations and gaseous friction.
These previous studies allowed us to identify and quantitatively study the onset
of orbital crossing, and in particular its progressive wave-like inward propagation,
as well as its damping by gas drag, for individual specific perturber configurations.
We here intend to take these studies a step further and extend them to the
general case of any given binary system configuration.
In a first step, we investigate the purely gravitational problem which is the basis
to understand more advanced results including gas drag. For this gas-free system,
our numerical code is used to empirically
derive analytical estimates of the orbital crossing location
and of the induced encounter velocities increase.
These general laws are expressed as a function of the
free parameters here explored: the binary separation $a_{b}$, the companion's
eccentricity $e_{b}$ and its mass $m_{b}$. We pay particular attention
to the timescale for the inward propagation of the orbital crossing ``wave''.
The second part of our work is devoted to the study of how gas drag
affects these pure gravitation induced results.
Frictional drag by the gas of the protoplanetary disk might indeed
play an important role here. If efficient enough, it can restore the periastron
alignment \citep[e.g.][]{mascho98},  
preventing orbital crossing of orbits with different semimajor  
axes. At the same time, it partially damps the amplitude of  
oscillations in eccentricity induced by the companion star. Another important
aspect of gas drag is that it is size dependent, and thus introduces differential
effects between bodies of different radius \citep[e.g.][]{mascho00,kort01,the04}.
Semi-empirical analytical laws can here no longer be derived and we have to rely
on full scale numerical simulations including the gaseous friction force.
We apply a simple gas drag model, where the gas disk rotates on circular orbits
with non-Keplerian velocities due to a pressure gradient.
We pay special particular attention to the crucial problem of encounter velocity
evolution between objects of different sizes triggered by the differential
orbital phasing induced by gas drag.

In order to interpret our encounter velocity distributions in terms
of accreting or erosive impacts, we depart from
the simplified approximation of \citet{whit98} that all encounter
velocities higher than $\Delta v_{lim}=$100m.s$^{-1}$ should
lead to catastrophic disruption.
The underlying assumption that the limiting velocity for disruption
is independent of planetesimal masses might indeed be an
oversimplification, even for non--gravitationally bound bodies \citep[e.g.][]{benz99}.
Furthermore, the accretion/disruption dichotomy overlooks the fact that,
in many cases, $\Delta v$ increase stops accretion not by triggering fully
catastrophic disruption but rather by erosive cratering, and that erosion is
in fact often
the dominant mass removal mechanism in dynamically excited systems 
\citep[see for instance the quantitative study of][]{theb03}. Another
important point is to consider $\Delta v_{lim}$ values for impacts
between objects of different sizes, firstly because runaway accreting bodies
are believed to grow mostly through accretion of field planetesimals
significantly smaller than themselves, and secondly because any ``real'' population
of planetesimals necessarily has a statistical spread in object sizes.
For all these reasons, we here make use of detailed
statistical models of collision outcomes for estimating if impacts are
in the accreting or eroding regimes.

\section{THE GRAVITATIONAL PROBLEM}

We consider the case where the planetesimal disk and the binary
orbital plane are coplanar.
We consider a system of bodies initially on
unperturbed circular orbits $e=0$, which is the usual assumption
for such perturbation studies \citep[see for example][]{hep78,whit98}.
This assumption is implicitly equivalent to
considering that all planetesimals begin to ``feel'' the secondary
star perturbations, i.e. that they become big enough to decouple from the
surrounding gas, at the same time $t_{0}$, or more exactly that
any possible spread in $\Delta t_{0}$ is negligible compared to the
typical timescales $t_{0_{sec}}$ and $t_{0_{acc}}$
of both secular perturbations and runaway growth.
Values for $\Delta t_{0}$ depend on
the planetesimal formation process, and in particular on its timescale.
This crucial issue is out of the scope of the present study, but
it is important to stress that it is far from having been solved yet.
There is indeed still intense debate going on between supporters of
the two main concurrent models, i.e. the gravitational instability
scenario \citep[e.g.][]{gold73,you02,you04},
where kilometre--sized bodies directly form from small solids in dense
unstable solid grain layers, and the collisional-chemical sticking
scenario \citep[e.g.][]{wei80,dom97,dul05}, in which planetesimals are the
results of progressive mutual grain sticking. 
In the gravitational instability model, the
planetesimal formation timescale is likely to be negligible compared to the
typical timescales for runaway growth and companion star perturbations. In
this case, the moment when planetesimal begin to feel the secular
perturbations would more or less coincide with the start of
runaway accretion and initial coapsidality between all bodies
would probably be a valid assumption.
Things are less clear in the sticking scenario, where there is no
abrupt ``leapfrog'' from grains to planetesimals. The decoupling
from the gas is more progressive, but some simulation
results suggest that the whole growth process from grains to kilometre-sized
bodies might not exceed a few $10^{3}$ years \citep{wei00}, i.e. still
shorter than both runaway growth and secular perturbations timescales.
However, the current
understanding of both these scenarios is still limited, each of them
having major physical obstacles to overcome, i.e. the
formation of a thin dense layer of solids in the instability scenario
and a way of resisting 50m.s$^{-1}$ impacts in the sticking model
\citep[e.g.][]{you04} so that it is very difficult to give any realistic
estimate for $\Delta t_0$.
We thus believe the present set of initial
conditions to be the most generic one in the current state of the
dust-to-planetesimals formation process knowledge.
For a critical discussion on this issue, see section 5.1.

\subsection{Numerical example}

\begin{figure}
\includegraphics*[width=1.\textwidth]{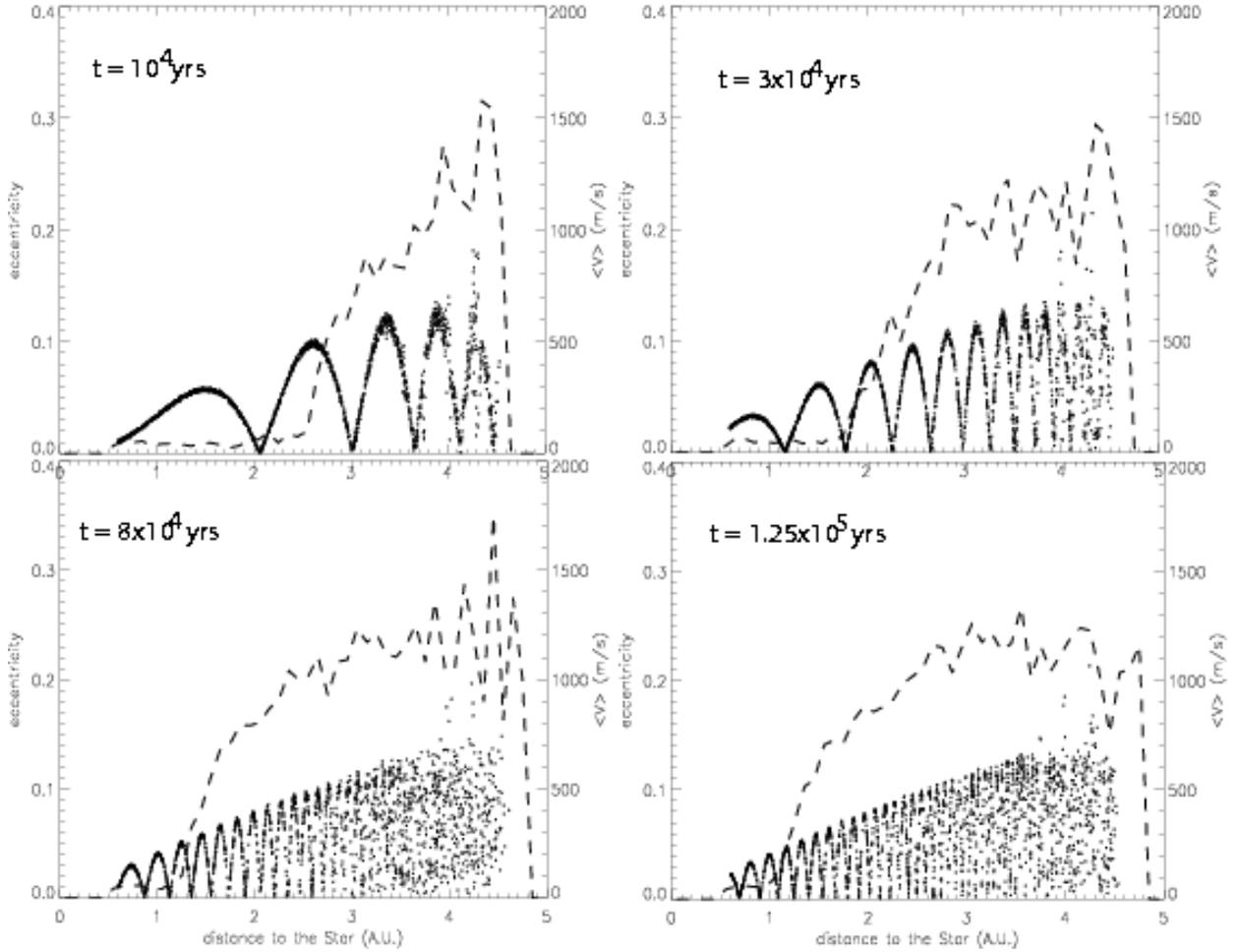}
\caption[]{Evolution of a test particle population perturbed by a stellar
companion with $m_{b}=0.25$, $a_{b}=20\,$AU and $e_{b}=0.3$. The dotted line
represents the distribution of average encounter velocities within the system.
}
\label{numex}
\end{figure}

Our code is an updated version of a program initially developed for the
study of planetesimal systems perturbed by a giant planet
\citep[e.g.][]{the02} and recently used for the study of the $\gamma$
Cephei system (TH04). Let us here briefly outline that it deterministically
follows the dynamical evolution of a population of massless test particles and
that it has a close encounter search algorithm which enables tracking of all
mutual encounters as well as accurate estimations of relative velocities
at impact.
Fig.\ref{numex} shows a typical outcome for such a numerical run.
The dynamical behaviour of the test particle system is dominated
by eccentricity oscillations forced by the
companion star. Once they begin to develop,
the amplitude of these oscillations is almost constant with time and
increases with proximity to the companion (i.e. in the outer regions).
The frequencies of these oscillations do also increase with
proximity to the companion but they do not remain constant and
continuously increase with time. 
It is important to notice that, at least in the beginning, these
large eccentricity oscillations do not lead to high encounter velocities
because of the strong phasing between neighboring orbits.
As times goes by, however, these oscillations get narrower and
narrower, so that at some point the phasing can no longer prevent
orbit crossing. This leads to a sudden increase of encounter velocities
which may reach very high values, typically of the order of $10^{3}$m.s$^{-1}$.
These values are high enough
to prevent the accretion of any kilometre--sized planetesimal.
As a consequence, the planetesimal accretion should be stopped in
the orbit-crossing region, provided that no bigger object had the time to form
before the orbital-crossing occurs.
The transition between
non orbit-crossing and orbit-crossing regions is very sharp, and
one might consider that it occurs at one given semi-major axis $a_{cross}$.
Notice that $a_{cross}$ progresses inward, so that the
region of low $\Delta v$ gets narrower and narrower with time.

\subsection{Analytical derivation}

In order to avoid performing a tremendous amount of numerical runs
exploring the full space of free parameters, we examine
the possibility to get a reasonably accurate analytical formula
giving the value of $a_{cross}$.
Since we intend to study how accretion around a solar type primary
star is affected by the companion's perturbations, we chose
a system of units related to the primary star.
All masses are thus renormalized to the mass of the
primary ($1\,M_{\odot}$), all distances to 1\,AU and all times
to 1\,yr, i.e. the orbital period at 1\,AU from the primary.

\subsubsection{Revised expression for the secular approximation}

The usual way to analytically describe perturbations triggered by
a companion is to use the simplified secular perturbation theory
equations for the eccentricity and longitude of periastron
\citep[e.g.][]{hep78,whit98}:
\begin{equation}
e = \frac {5}{2} \, \frac{a}{a_{b}} \, \frac {e_{b}}{1-e_{b}^{2}}
\vert \sin \left( \frac {u}{2} t \right) \vert = e_{max}\,
\vert \sin \left( \frac {u}{2} t \right) \vert
\label{eosc}
\end{equation}
\begin{equation}
\tan(\varpi t) = - \frac {\sin ut}{1-\cos ut} ,
\label{wosc}
\end{equation}

where 
\begin{equation}
u = \frac {3}{2} \pi \frac {1}{\left(1-e_{b}^{2}\right)^{3/2}}
m_{b}  \frac {a^{3/2}}{a_{b}^{3}}.
\label{uequ}
\end{equation}

Equation \ref{eosc} gives a reasonably accurate estimation of the
amplitude of the eccentricity estimations, the relative error being
always smaller than $10 \%$ for the set of parameters explored hereafter.
However, due to the low order of its expansion,
the expression for the frequency $u$ of these oscillations 
is a poor match to the ones numerically obtained, and
differences can be as large as $70 \%$.
In order to empirically derive a more accurate prescription for $u$, we
performed several numerical test runs and get the revised expression
\begin{equation}
u = \frac {3}{2} \pi \frac {1}{\left(1-e_{b}^{2}\right)^{3/2}}
m_{b}  \frac {a^{3/2}}{a_{b}^{3}}
\, \left[ 1 + 32 \frac {m_{b}} {\left( 1-e_{b}^{2}\right)^{3} }
 \left( \frac{a}{a_{b}} \right)^{2} \right].
\label{realu}
\end{equation}

Equation \ref{realu} proved to be accurate within $5\%$ for the range of
parameters here explored.

\subsubsection{Orbital crossing location}

Directly applying the classical orbital crossing
criterion \citep[e.g. Equ.6 of][]{whit98} to Equs.\ref{eosc} and \ref{wosc}
results in an expression too complex 
to derive an analytical solution for $a_{cross}$.
However, we empirically observed from our test runs that orbital crossing
first occurs, within a given e-oscillation ``wave'', when particles of
high eccentricity within this wave are able to cross the orbit
of particles at the ``node'' (i.e. e$\simeq 0$) of the wave.
Furthermore, this crossing of the e=0 orbit occurs at almost the same
time for all particles within one given wave.
This crossing criterion might be written as
\begin{equation}
(a_0+\Delta a) \left(1-e_{(a_{0}+\Delta a)}\right) =  a_0 \Longleftrightarrow
(a_0+\Delta a)\,e_{(a_{0}+\Delta a)} = \Delta a,
\label{cross1}
\end{equation}
where $a_{0}$ is the semi-major axis of particles with $e=0$.
If we restrict ourselves to the region where the width of an e-oscillation is
always much smaller than its distance to the star (i.e. $\Delta a/a_0 <<1$), then
\begin{equation}
a_0\,e_{(a_{0}+\Delta a)} \simeq \Delta a \Longrightarrow
a_0 \left( \frac{\partial e}{\partial a} \right) _{(a_0)} \simeq  1 .
\label{cross2}
\end{equation}

Rewriting Equ.\ref{eosc} in the form $e_{max}\,sin(\theta_{(a)})$, it follows that
\begin{equation}
\left( \frac{\partial e}{\partial a} \right)_{(a_0)} =
e_{max}\left( \frac{\partial \theta}{\partial a} \right)_{(a_0)}
cos(\theta_{(a_0)}) = e_{max}
\left(\frac{\partial \theta}{\partial a}\right)_{(a_0)} \simeq \,
\frac{\pi}{2} \frac{e_{max}}{\Delta a_{m}},
\label{cross3}
\end{equation}
where $e_{max}$ is the maximum value of the oscillation wave (given by
Equ.\ref{eosc}) reached for
$a=a_{(e_{max})}$, and $\Delta a_{m}=a_{(e_{max})}-a_{0}$
is such that
\begin{equation}
\frac{u_{(a_{0}+\Delta a_{m})}}{2}t-\frac{u_{(a_{0})}}{2}t = \frac{\pi}{2}\,\,
\,\Longrightarrow\,\,\,{\Delta a_{m}}\frac{\partial u}{\partial a}=\frac{\pi}{t} .
\label{deltam}
\end{equation}
Equation \ref{realu} then gives
\begin{equation}
\Delta a_{m} = \frac{2\pi}{3K}\frac{1}{a^{1/2}
\left[1+\frac{224}{3}\frac{m_{b}} {\left( 1-e_{b}^{2}\right)^{3} }
\left(\frac{a}{a_{b}}\right)^{2}\right]}\,t^{-1} ,
\label{deltaa}
\end{equation}
where
\begin{equation}
K = \frac {3}{2} \pi \frac {1}{\left(1-e_{b}^{2}\right)^{3/2}}
m_{b}  \frac {1}{a_{b}^{3}} .
\label{K}
\end{equation}

\begin{figure}
\makebox[\textwidth]{
\includegraphics[width=.5\columnwidth]{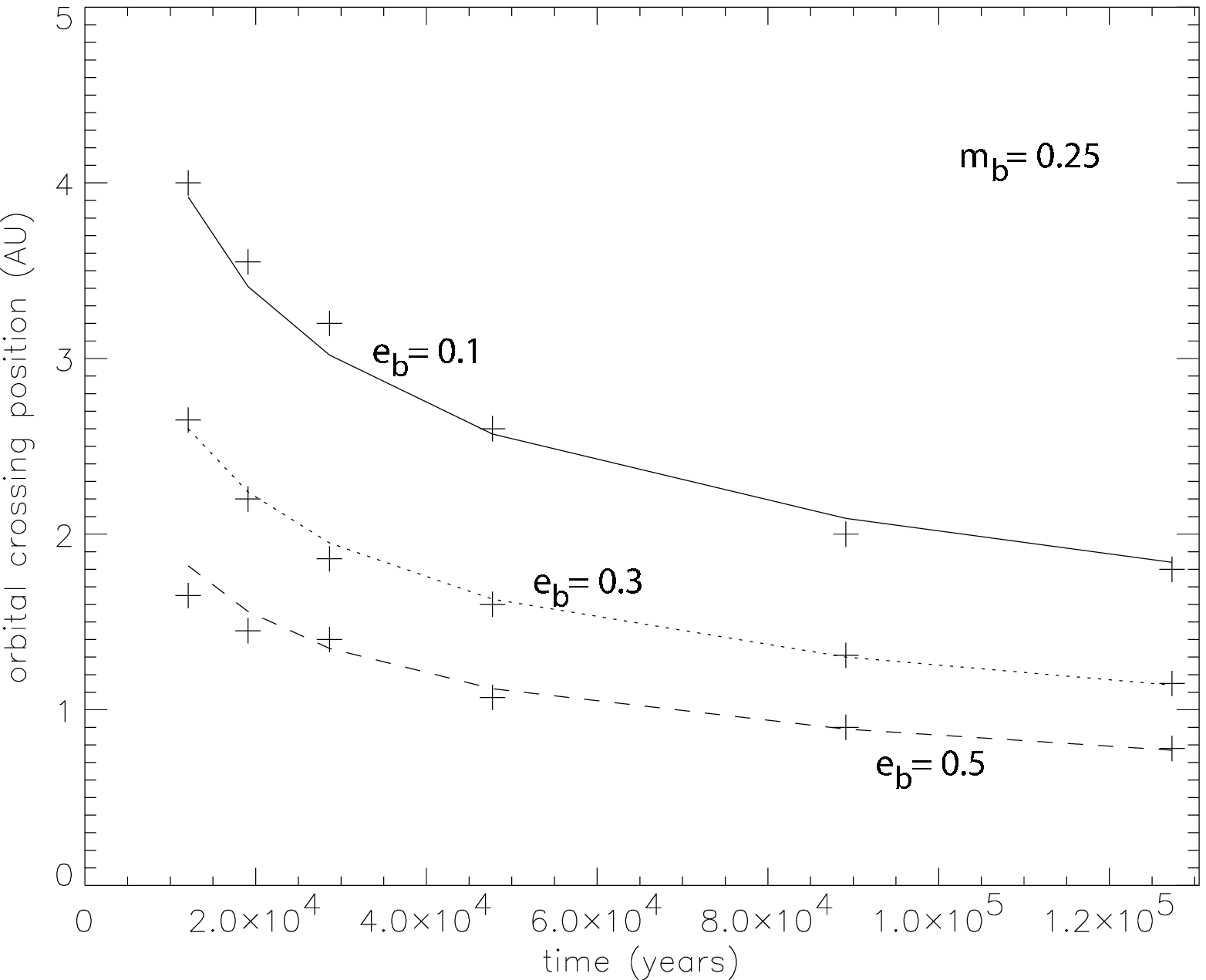}
\hfil
\includegraphics[width=.5\columnwidth]{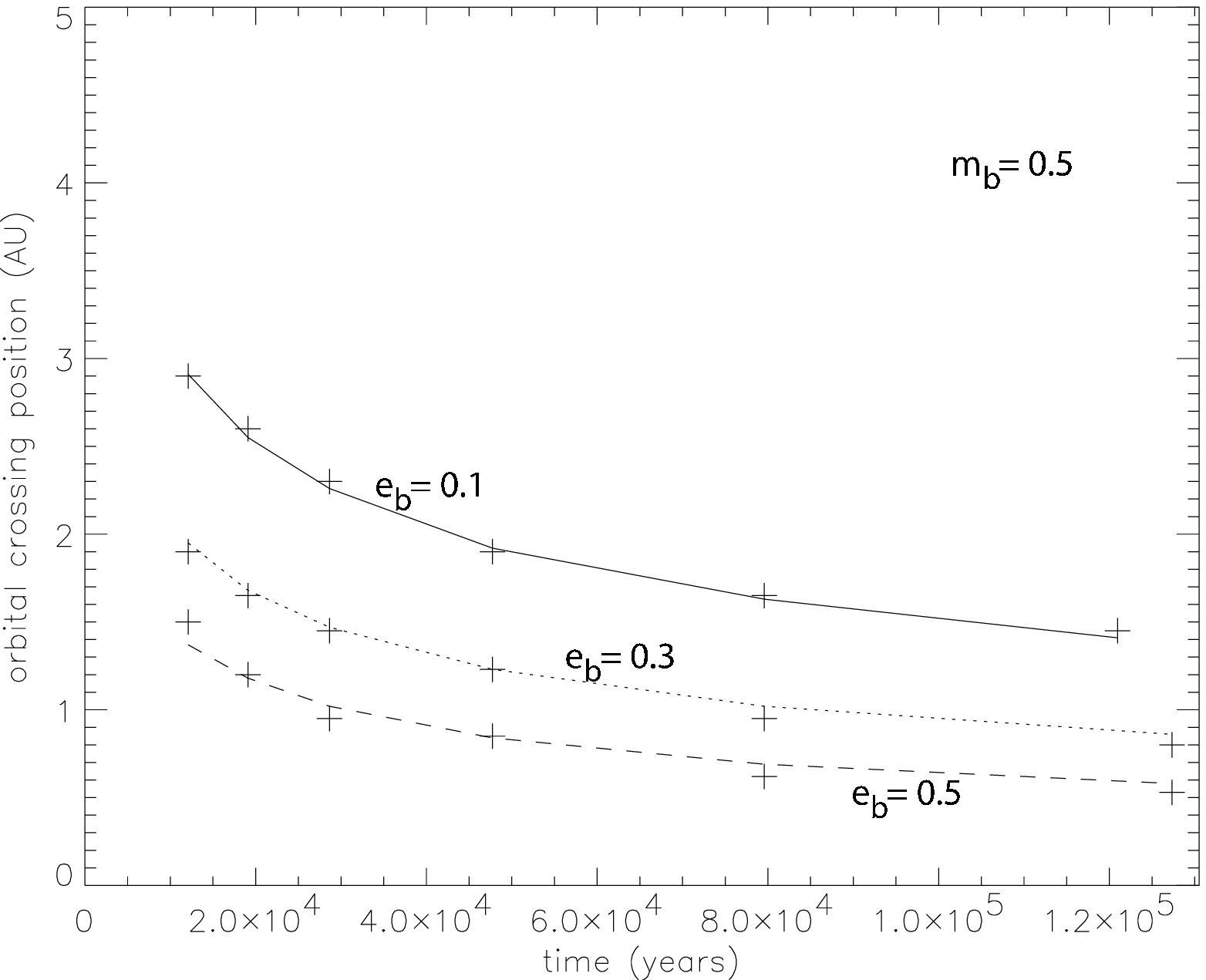}
}
\caption[]{Comparison between the analytical expression for orbital crossing
location given by Equ.\ref{realeq} (lines) and values obtained for
numerical test runs (crosses). The companion star has a semi-major axis
$a_{b}=20\,$AU and 3 different eccentricities.
Companion masses are given in solar mass units.
}
\label{comp}
\end{figure}

Putting Equ.\ref{eosc},  Equ.\ref{cross3} and Equ.\ref{deltaa}
into Equ.\ref{cross2} gives
\begin{equation}
C_{1}a^{1/2} \left[ \left(\frac{a}{a_{b}}\right)^{2}+
C_{2}\left(\frac{a}{a_{b}}\right)^{4} \right] = \frac{1}{t} ,
\label{realeq}
\end{equation}
where
\begin{equation}
C_{1} = \frac {45\pi}{16}  \frac {e_{b}}{\left(1-e_{b}^{2}\right)^{5/2}}
m_{b}  \frac {1}{a_{b}^{2}}
\,\,\, ;
C_{2} = \frac{224}{3} \frac {m_{b}} {\left( 1-e_{b}^{2}\right)^{3} } .
\label{C12}
\end{equation}
Equation \ref{realeq} is easily numerically solved and gives the value of
the orbital crossing location $a_{cross}$ for any given set of parameters.
The validity of Equ.\ref{realeq} has been checked by comparing its
solution to numerically obtained estimates for extreme values of the
free parameters of the problem ($a_{b}$, $m_{b}$, $e_{b}$).
As Fig.\ref{comp} clearly shows, there is always less than a $10\%$ relative
error between the analytical estimate and the numerical values.

\subsubsection{Relative velocities}

Another important parameter is the average values of relative velocities 
$\langle\Delta v \rangle$ in the orbital crossing region beyond $a_{cross}$.
From our series of test numerical runs, we have derived the simplified
empirical relation

\begin{equation}
\langle \Delta v \rangle _{(a)} \simeq 0.5\,e_{max(a)}\,v_{kep(a)},
\label{dv}
\end{equation}

which proved to be of reasonable accuracy, i.e. less than 10\% relative error.
Note that this formula gives a value which is smaller, by roughly a factor 2, than
the standard formulae giving average encounter velocities
for completely randomized orbits
$\langle \Delta v \rangle = (5/4)^{1/2}\,e\,v_{kep}$ 
\citep[e.g. Equ.17 of][]{lisste93}.
The main reason for this is that,
despite orbital crossing, orbits are not fully randomized.
A direct consequence of Equ.\ref{dv} is that
$\langle\Delta v \rangle$ has the same parameter
dependency as $e$ (it is in particular independent of the
companion mass $m_{b}$).

\subsubsection{Parameter space exploration}

We explore Equ.\ref{realeq} for a wide range of the companion's
orbital parameters and mass in order to map the region of parameter space
allowing or preventing accretion at a given distance of the primary
star. We consider as reference values for $a_{b}$, $e_{b}$
and $m_{b}$, the median values derived from \citet{duq91} for their
sample of G star binaries, i.e. $a_{b}\simeq 36\,$AU, $e_{b}\simeq 0.5$
and $m_{b}\simeq0.5$. These parameters are then individually explored
(Figs.\ref{ebab}, \ref{mbab} and \ref{tlab}).
We focus here on the case of planets in the habitable zone and thus
take 1\,AU as a reference region of interest for the orbital crossing
location.

One additional crucial parameter is the time $t_{cr}$ at which orbital
crossing occurs. We will here pay particular attention to
values between $5\times10^{3}$ and $10^{5}$\,years, i.e.
a conservative range for typical runaway growth timescales.
Figure \ref{tlab} clearly illustrates the time dependency,
showing a power law dependency between $t_{cr}$ and $a_{b}$
for $t_{cr}$ greater than a few thousand years.
By thoroughly exploring the influence of all parameters
on $t_{cr}$, we were able to derive the following empirical law,
valid in the $a_b>10\,$AU and $0.05<e_b<0.8$ parameter range:

\begin{equation}
t_{cr}=\\
6.98\times10^{2} 
\frac{\left(1-e_{b}^{2}\right)^{3}}{e_{b}}
\left(\frac{m_{b}}{1M_{\odot}}\right)^{-1.1}
\left(\frac{a_{b}}{10\rm{AU}}\right)^{4.3}
\left(\frac{a_{cross}}{1\rm{AU}}\right)^{-2.8}
\rm{yrs}.
\label{tcr}
\end{equation}

This gives in turn the location of $a_{cross}$ as a function of time
and companion star parameters:

\begin{equation}
a_{cross}=0.37
\frac{\left(1-e_{b}^{2}\right)^{1.07}}{e_{b}^{0.36}}
\left(\frac{m_{b}}{1M_{\odot}}\right)^{-0.39}
\left(\frac{a_{b}}{10\rm{AU}}\right)^{1.53}
\left(\frac{t}{10^{4}yr}\right)^{-0.36}
\,\rm{AU} \,.
\label{acr}
\end{equation}

This formulae, contrary to Equ.\ref{realeq}, has been
purely empirically obtained instead of analytically derived.
As for Equ.\ref{realeq}, we have checked its validity by comparison
with a set of numerical tests exploring all extreme values for the
parameters here explored. These comparisons showed that the value
given by Equ.\ref{acr} was always within a satisfying
$10\%$ of the numerical result. 

The corresponding expressions for 
the critical values of the companion's semi-major axis
and mass (the expression for $e_{b.cr}$ is more 
complicated and has to be numerically solved)
as a function of $a_{cross}$ and time are then easely derived and read

\begin{equation}
a_{b.cr}=18.1
\frac{e_{b}^{0.23}}{\left(1-e_{b}^{2}\right)^{0.69}}
\left(\frac{m_{b}}{1M_{\odot}}\right)^{0.26}
\left(\frac{a_{cross}}{1\rm{AU}}\right)^{0.65}
\left(\frac{t}{10^{4}yr}\right)^{0.23}
\,\rm{AU}
\label{ab}
\end{equation}
\begin{equation}
m_{b.cr}=8.95\times10^{-2}
\frac{\left(1-e_{b}^{2}\right)^{2.72}}{e_{b}^{0.91}}
\left(\frac{a_{b}}{10\rm{AU}}\right)^{3.90}
\left(\frac{a_{cross}}{1\rm{AU}}\right)^{-2.54}
\left(\frac{t}{10^{4}yr}\right)^{-0.91}
\,M_{\odot}.
\label{mb}
\end{equation}

\section{EFFECT OF GAS DRAG}

The results displayed in the previous section have been obtained
for a simplified system where gravitational perturbing effects are the only forces
acting on the planetesimal population. In the current standard
planetary formation scenario however, the initial stages of
planetesimal accretion are believed to take place in the presence
of significant amounts of primordial gas. 
Gas drag affects planetesimal orbits by partially damping their
eccentricity as well as forcing periastron alignment among bodies
of the same size \citep[e.g.][]{mascho98}, thus
cancelling the secular orbital oscillations induced by the perturber.
These two effects might have drastic consequences on the onset of orbital crossing
and on the values of impact velocities.
In his earlier study, \citet{whit98} claimed the neglect of
gas drag to be justified for bodies of size $\ge\,$10km. However, this
argument might not hold for 2 reasons: 1) It is not clear if the initial
planetesimal sizes were $\ge\,$10km (simply because it is not clear what
this initial size actually is), and smaller objects should be
more significantly affected by gas drag (this force being $\propto1/R$).
2) Even for large planetesimals, the neglect of gas drag for $\ge\,$10km
objects is a result that holds only for unperturbed planetesimal populations
with low eccentricities. In the present case, we are dealing with high $e$
orbits for which gaseous friction is more effective, since
gaseous eccentricity damping is expected to be $\propto e^{2}$ \citep{ada76}.
A study of planetesimal accretion in binary systems can thus probably
not dispense from addressing the crucial issue of gas drag
effects. The non-conservative characteristics of gaseous friction as well
as the number of additional free parameters it introduces, i.e.
gas density and distribution, planetesimal absolute and relative physical
sizes, etc.., renders a semi-empirical analytical analysis in
the spirit of the one presented in the previous section almost impossible.
We have here to rely on numerical simulations.
Numerical studies of gaseous friction have been undertaken by the
present team in several previous works for specific perturbed systems,
e.g. \citet{mascho98,mascho00,the02} and TH04 for 
the specific case of the $\gamma$ Cephei binary system.
The aim of the present study is to generalize these previous
studies to the general case of any binary system.

\subsection{Modeling}

The numerical code is the same as the one used in \citet{the02} and TH04.
We follow \citet{WD85} and model the gas force as
\begin{eqnarray}  
        \vec{F} & = & - K v \vec{v}  , 
\end{eqnarray}  
where $\vec{F}$ is the force per unit mass, $\vec{v}$ the velocity  
of the planetesimal with respect to the gas, $v$ the velocity modulus,  
and $K$ is the drag parameter. It is a function of the physical  
parameter of the system and is defined as:  
\begin{eqnarray}  
        K & = & \frac{3 \rho_{\rm g} C_d} {8 \rho_{{\rm pl}} R_{}}  ,
\end{eqnarray}  
where $\rho_{\rm g}$ is the gas density, $\rho_{{\rm pl}}$ and $R$ the  
planetesimal density and radius, respectively. $C_d$ is a  
dimensionless coefficient related to the shape of the body  
($\simeq 0.4$ for spherical bodies).  

Exploring the gas drag density profile as a free parameter would
be too CPU time consuming and we shall restrict ourselves to one
$\rho_{\rm g}$ distribution. We
assume the standard Minimum Mass Solar Nebula (MMSN) of \citet{haya81},
with $\rho_{\rm g}=\rho_{\rm g0}(a/1AU)^{-2.75}$ and
$\rho_{\rm g0}=1.4\times10^{-9}$g.cm$^{-3}$. 
We take a typical value $\rho_{\rm pl}=3\,$g.cm$^{-3}$.
Since the initial sizes of accreting planetesimals are not very well constrained,
we shall consider two types of detailed simulations, i.e.
one for a typical ``small planetesimals'' case, where
$1<R<10\,$km, and one for a typical ``big planetesimals'' case with
$10<R<50\,$km. For each case, the main outputs are the values
of encounter velocities $\Delta v_{(R1,R2)}$ for all possible target-projectile
pairs of sizes $R_1$ and $R_2$.

To interpret these results in terms of accreting or eroding
impacts, the obtained $\Delta v_{(R1,R2)}$ values have to be compared
to reference semiempirical models of collisional outcomes. The core assumption
for these models are the prescriptions for cratering excavation factors
and the disruption threshold parameter $Q_*$. Several different
such prescriptions, based on laboratory experiments and/or
energy scaling considerations, are currently available. One has however
to remain careful since many of these prescriptions are far
from agreeing with each other
and often predict very different physical outcomes for a given set of
$R_1$, $R_2$ and $\Delta v_{(R1,R2)}$ parameters \citep[a striking illustration of this
can be found in Fig.8 of][where values of $Q_*$ from different authors
are compared]{benz99}.
We shall here analyze our $\Delta v_{(R1,R2)}$ values using
the statistical model of \citet{theb03}, a detailed numerical tool developed
for the study of collisional cascades in extrasolar disks, which can accommodate
for different $Q_*$ and cratering excavation coefficient prescriptions.
We here explore 3 different $Q_*$ prescriptions, i.e. \citet{mar95},
\citet{hol94} and \citet{benz99} and two ``soft'' and ``hard material''
cases for the cratering excavation coefficient.
We remain as careful as possible and shall only consider that impacts are
preferentially accreting or eroding when all tested prescriptions agree, and
do not derive definitive conclusions for the intermediate ``limbo''
region where different outcomes are obtained depending on the 
assumed collisional formalism.

To get statistically significant results all runs include $N=10^{4}$ test particles.

\subsection{Typical behaviour}

Let us first present results obtained for two representative ``pedagogical'' 
companion configurations: 1) $a_b = 10\,$AU, $e_b = 0.3$, $m_b = 0.5$,
a highly perturbed case for which Equ.\ref{tcr} predicts fast orbital
crossing, at 1\,AU, under pure secular perturbations
2) $a_b = 20\,$AU, $e_b = 0.4$, $m_b = 0.5$, a less perturbed case
with no predicted orbital crossing occurring before $5\times10^{4}$yrs.
For these two illustrative examples, we display the results in
full matrices of average encounter velocities $\Delta v_{(R1,R2)}$
for all impacting pairs of sizes $R_1$ and $R_2$ in the
``small'' and ``big'' planetesimals runs.
Presented $\Delta v$ values are averaged over a typical runaway growth
time interval $0<t<2\times10^{4}$yrs, but no crucial time dependent information is
lost, since in general $\Delta v_{(R1,R2)}$ values set in on timescales of only a few
$10^{3}$yrs.

\subsubsection{highly perturbed case: $a_b = 10\,$AU, $e_b = 0.3$, $m_b = 0.5$}

Fig.\ref{exsmall} shows the typical dynamical evolution
for two planetesimal populations of different sizes. The most obvious feature
is the forced orbital alignment among equal-sized bodies. In the present
example, it is strong enough to prevent any orbital crossing, within
$2\times10^{4}$years, for all {\it equal-sized} small planetesimals (Fig.\ref{vitevol}).
For bigger objects in the $50$\,km size range, orbital crossing occurs,
at approximately the time predicted by the previous section's set of equations,
although for encounter velocities significantly smaller than in a gas free case.
For bodies of {\it different} sizes however, results are almost the exact opposite.
Indeed, orbital alignment strongly depends on planetesimal radius:
smaller objects tend to align more quickly and
towards periastron values located at $\simeq$ 270 degrees from that of the perturber
$\varpi_p$, while bigger bodies need more time to align their orbits and do it towards
periastron values closer to $\varpi - \varpi_p \simeq 360^{o}$
\citep[a result already identified by][]{mascho98}. This is clearly illustrated
in Fig.\ref{exsmall}, with 1km objects on fully phased orbits after
$3\times10^{3}\,$yrs while 5km bodies haven't reached complete
alignment yet, especially in the outer regions where
residual secular oscillations are still visible. The consequence of this
differential periastron alignment is a fast and significant increase
of encounter velocities between these 2 populations (Fig.\ref{vitevol}).
The full $\Delta v_{(R1,R2)}$ matrixes of Tabs.\ref{dvmatrix1a} and \ref{dvmatrix1b}
show that this result holds for all sizes: differential phasing always
results in $\Delta v$ increase as soon as $R_1 \neq R_2$. 
In contrast, the $\Delta v$ damping effect between equal-sized objects appears as only
a marginal feature, especially in the 1-10km range. Indeed, if as a
first approximation we parameterize the collisional efficiency of a $\Delta v_{(R1,R2)}$
impact by the amount of kinetic energy $Ec_{(R1,R2)}$ delivered to the target $R_2$
by an $R_1$ ($<R_2$) impactor, then an exploration of the values displayed in
Tab.\ref{dvmatrix1a} shows that
$Ec_{(R1,R2)}$ always exceeds $Ec_{(R2,R2)}$ for $R_2 \leq 10\,$km.
As a matter of fact, for most impacting pairs in this size range
the delivered kinetic energy peaks at roughly $R_1 \simeq 1/2R_2$. 
The crucial result is in any case that, for almost all $R_1 \neq R_2$ impacts
of the small planetesimals run, encounter velocities exceed by far
the limit for accretion and the system should clearly undergo erosion
(Tab.\ref{dvmatrix1a}).

Things are less simple in the big planetesimals run. The only low velocity
collisions are those between equal-sized bodies smaller than 20\,km. For bigger
objects gas drag can no longer prevent orbital crossing and
the corresponding velocity increase.
For objects of different sizes, differential orbital phasing has the same
velocity increase effect as in the small planetesimal case; although it is a bit
more complicated here, where the obtained $\Delta v_{(R1,R2)}$ is often a combination
of an orbital crossing and of a differential phasing term (for which
the delivered kinetic energy $Ec_{(R1,R2)}$
peaks for $R_1/R_2$ values close to $\simeq 1/3$).
Note however that
encounter velocities never reach values for which we might be sure that
erosion dominates over accretion. $\Delta v_{(R1,R2)}$ values are here in
the ``limbo'' range where empirical collisional outcome models disagree on the
net accretion vs. erosion balance (Tab.\ref{dvmatrix1b}).

\subsubsection{moderately perturbed case: $a_b = 20\,$AU, $e_b = 0.4$, $m_b = 0.5$}

This case should in principle be radically different from the previous one, as
Equ.\ref{tcr} predicts that no orbital crossing should occur within the timeframe
of the simulation. For the small planetesimals run however,
results are remarkably similar to those of the 
$a_b = 10\,$AU, $e_b = 0.3$, $m_b = 0.5$ case:
differential orbital phasing induced by gas drag
restricts low velocity impacts to a narrow $R_1 \simeq R_2$ diagonal and
$\Delta v_{(R1,R2)}$ are clearly in the eroding regime for a large majority
of impacting pairs (Tab.\ref{dvmatrix2a}). As in the previous example,
$Ec_{(R1,R2)}$ always exceeds $Ec_{(R2,R2)}$ and
the delivered kinetic energy also peaks for $R_1 \simeq 1/2R_2$.
Differences are nevertheless observed for the $10<R<50\,$km run, where
although velocity increase is observed for almost all $R_1 \neq R_2$ pairs,
its amplitude remains more limited than for the previous highly perturbed case. 
The main reason for this difference is that here we have only the gas drag induced
velocity increase but no contribution due to secular orbital crossing.
This is why $\Delta v_{(R1,R2)}$ remain relatively small for the bigger,
less affected by gas drag, objects (in the $R\ga25$km range).
As a consequence, most of mutual collisions in the ``big planetesimals'' run
turn out to result in net accretion (Tab.\ref{dvmatrix2b}).

\subsubsection{radial drift}

In addition to orbital alignment and eccentricity damping, 
it is well known that gas
drag also forces inward drift of planetesimals \citep[e.g.][]{the04}.
This drift could in principle also be the source of a
$\Delta V$ increase: the radial drift
being size dependent it might bring to the same location
objects "carrying" with them different periastron values depending
on their region of origin.
However, this effect appears to be negligible here. 
Indeed, the average inward drift of a 1\,km object (the smallest
size considered here)
starting at 1\,AU is $\simeq 1.6\times10^{-5}$AU.yr$^{-1}$
for the highly perturbed case.
This means that it has drifted by less than 0.05\,AU 
by the time complete periastron alignment is reached, i.e.
$3.\times 10^{3}\,$years. From Fig.\ref{exsmall} it is easy to see
that the difference in periastron between 1.05 and 1\,AU 
is small compared to the $local$ periastron difference, at 1\,AU,
between a 1\,km and a 5\,km object. In other words, local periastron
phasing effects are much more efficient (i.e. fast) than periastron
"transport" from one place to another, even in the extreme
case where objects would carry their
periastron without modifications from one place to another.
Another proof of the insignificance of this effect is that
if it was efficient then it should also affect the $\Delta V$
distribution between $equal$--sized objects, since even same--size
objects originating from different regions have different
drift rates and could thus be brought together. But, as
already mentioned, no $\Delta V$ increase 
is observed in our simulations for equal--sized planetesimals
(except for the non-gas drag induced effects on bigger planetesimals).

\subsection{$a_b$ and $e_b$ parameter exploration}

For obvious computing time constraints, it is impossible to thoroughly
explore all companion orbital parameters with simulations as detailed as those
presented in the previous section. We shall narrow our
exploration range by fixing the companion's mass to a typical  value
$m_b=0.5$ that of the primary and explore $a_b$ and $e_b$ as free parameters
in the $10<a_b<50$AU and $0.05<e_b<0.9$ ranges. A total number of
126 runs have been performed.
As already mentioned, we perform, for each companion configuration,
2 runs, for a ``small'' and a ``big'' planetesimals case respectively.
For sake of readability of the results, we display the obtained
$\Delta v_{(R1,R2)}$ values for specific $R_1$ and $R_2$ values corresponding to
a typical pair of impacting bodies for each case.
We take $R_1=2.5$ and  $R_2=5\,$km as representative values
for the small planetesimals run since, for most of
the $e_b$ and $a_b$ range explored here, simulations confirm
the results obtained for the previous specific examples, i.e.
that bodies delivering the maximum kinetic energy 
to a $R_2$ target are roughly those of size $R_1 \simeq 1/2R_2$.
For the bigger planetesimals, the delivered kinetic energy peaks
for somewhat smaller $R_1/R_2$ ratios and we consider
$R_1=15$ and  $R_2=50\,$km as typical example values.
As in the previous section, we display encounter velocities averaged
over $0<t<2\times10^{4}$yrs.

\subsubsection{small planetesimals case}

A first important result is that we never observe velocity increase between
equal-sized small planetesimals, i.e.
gas drag orbital alignment prevents orbital crossing
from occurring for all the ($a_b, e_b$) space explored here.
However, we observe a generalization of the previous section's result, i.e.
a dramatic $\Delta v_{(R1,R2)}$ increase as soon as $R_1 \neq R_2$.
This is clearly illustrated in Fig.\ref{dv5-2.5} for a typical
$R_1=2.5$ and  $R_2=5\,$km pair. 
We can schematically divide this graph into 3 regions:
\begin{itemize} 
\item The region where encounter velocities remain at their initial low values
($\Delta v \leq 10\,$m.s$^{-1}$).
This fully undisturbed region is confined to a narrow strip close to $e_b=0$ except
for large $a_b > 40$AU values. Its extent is very limited
when compared to the corresponding undisturbed region in the gas free
case, i.e. the one given by the limit for orbital crossing at $t\simeq2\times10^{4}$yrs
(see Fig.\ref{ebab})
\item The $10\leq\Delta v \leq 100\,$m.s$^{-1}$ region. Here, encounter velocities
begin to be significantly increased by differential orbital phasing.
The net collisional outcome remains however uncertain, since we are in the
``limbo'' range where the balance between accretion and erosion depends
on the assumed collisional outcome prescription.
This intermediate zone covers a large
fraction of the ($a_b, e_b$) space. Interestingly, its borders are almost
independent of $e_b$ for $a_b<20$AU.
\item The $\Delta v \geq 100\,$m.s$^{-1}$ region where encounter velocities are
increased to values always exceeding the net erosion threshold.
This zone still covers an area much more extended than that of the orbital-crossing
region of the gas free case. This is particularly true for companion
semi-major axis comprised between 20 and 40\,AU.
\end{itemize}

\subsubsection{large planetesimals case}

\begin{figure} 
\includegraphics[angle=0,origin=br,width=1.00\columnwidth]{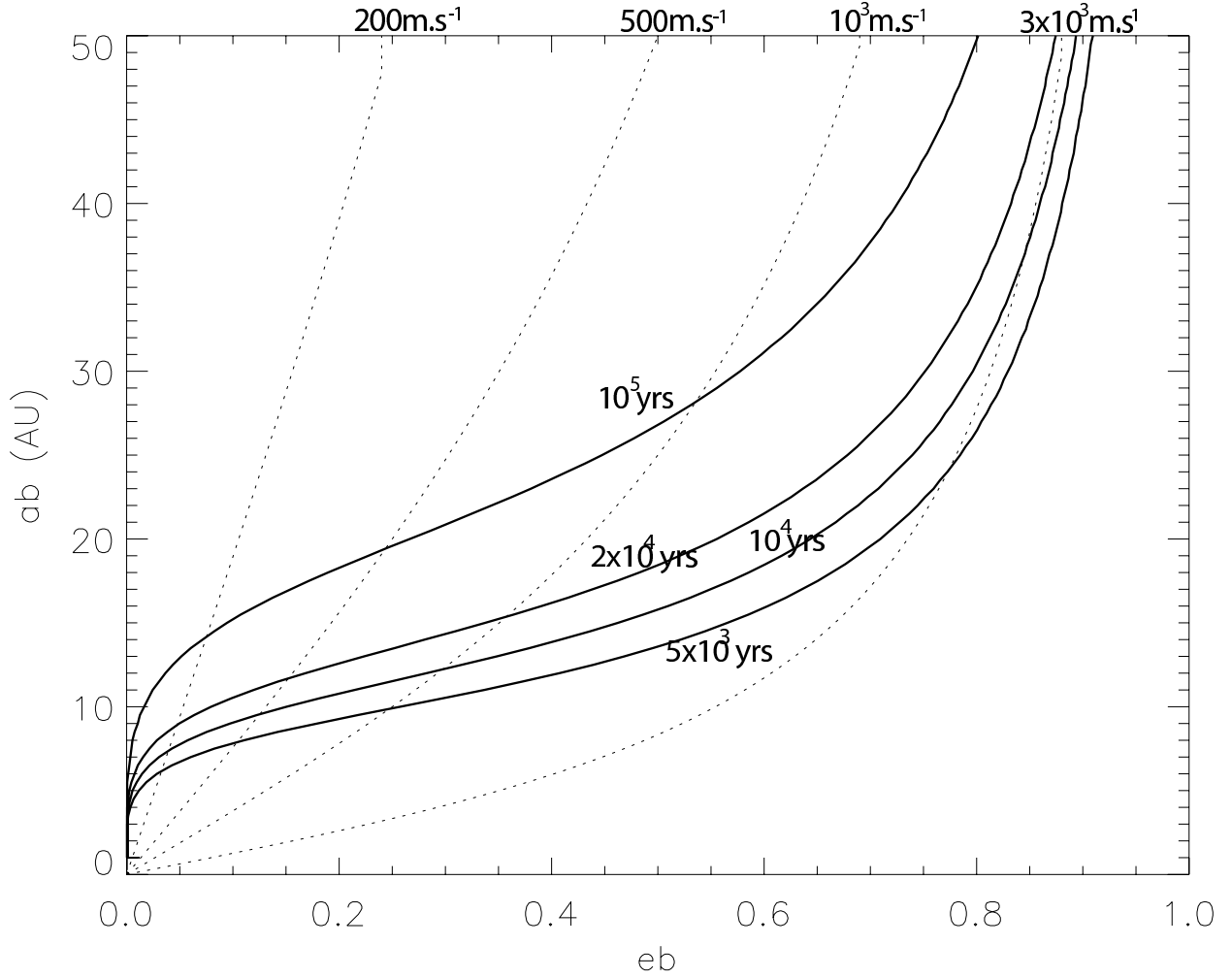}
\caption[]{Value of the minimum companion semi-major axis,
$a_{b.cr1}$, leading to orbital
crossing of planetesimals at 1 AU,
for different values of the crossing time $t_{cr}$,
as a function of the companion 
eccentricity. The companion's mass is fixed, with $m_{b}=0.5$.
The dotted lines represent constant values
of $\langle \Delta v \rangle$ at orbital crossing, as given by Equ.\ref{dv}
} 
\label{ebab} 
\end{figure}

\begin{figure} 
\includegraphics[angle=0,origin=br,width=1.00\columnwidth]{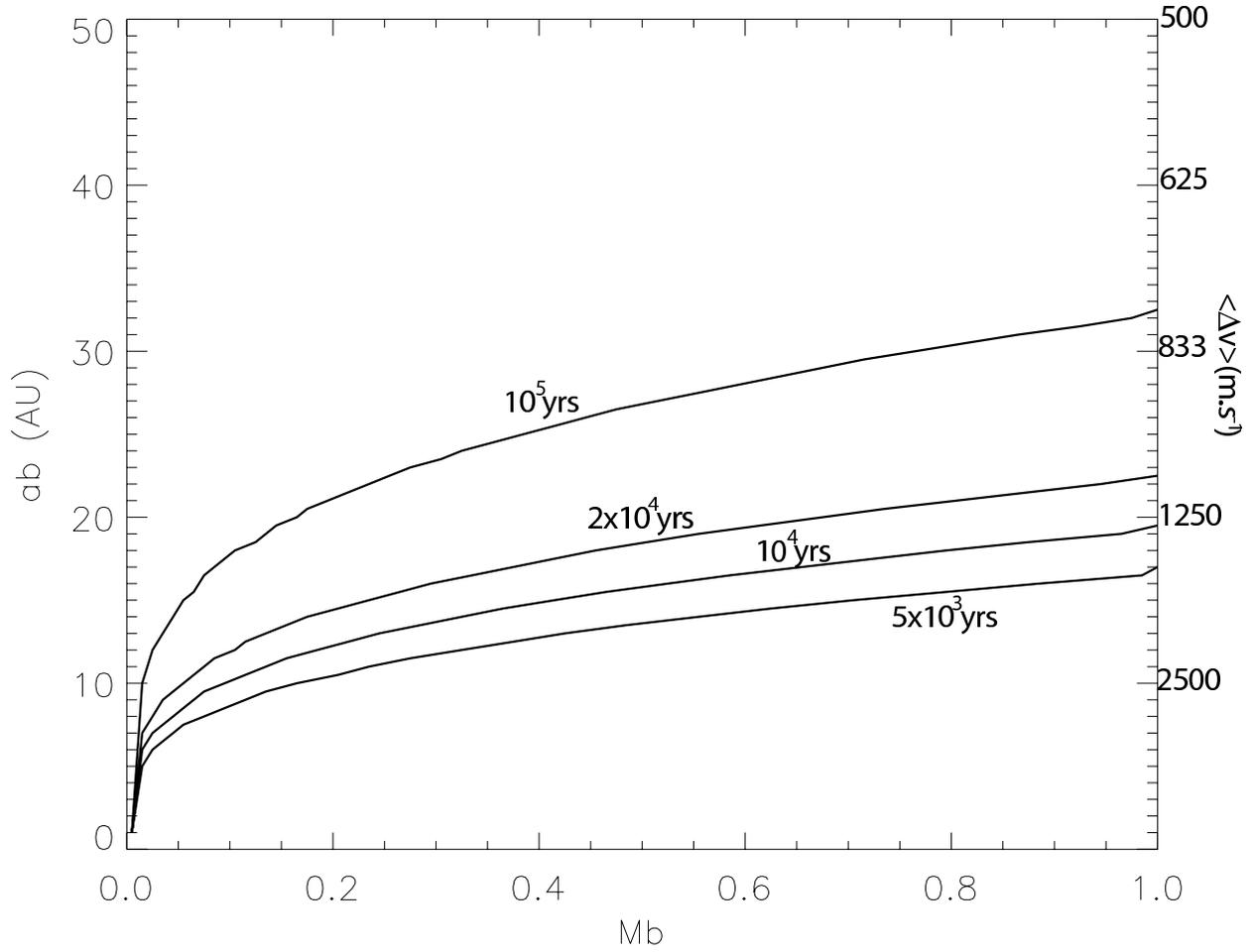}
\caption[]{Value of $a_{b.cr1}$ (same definition as in Fig.\ref{ebab}),
for the same 4 values of $t_{cr}$ as in Fig.\ref{ebab},
as a function of the companion 
mass. The companion's eccentricity is fixed, with $e_{b}=0.5$.} 
\label{mbab} 
\end{figure}

\begin{figure} 
\includegraphics[angle=0,origin=br,width=1.00\columnwidth]{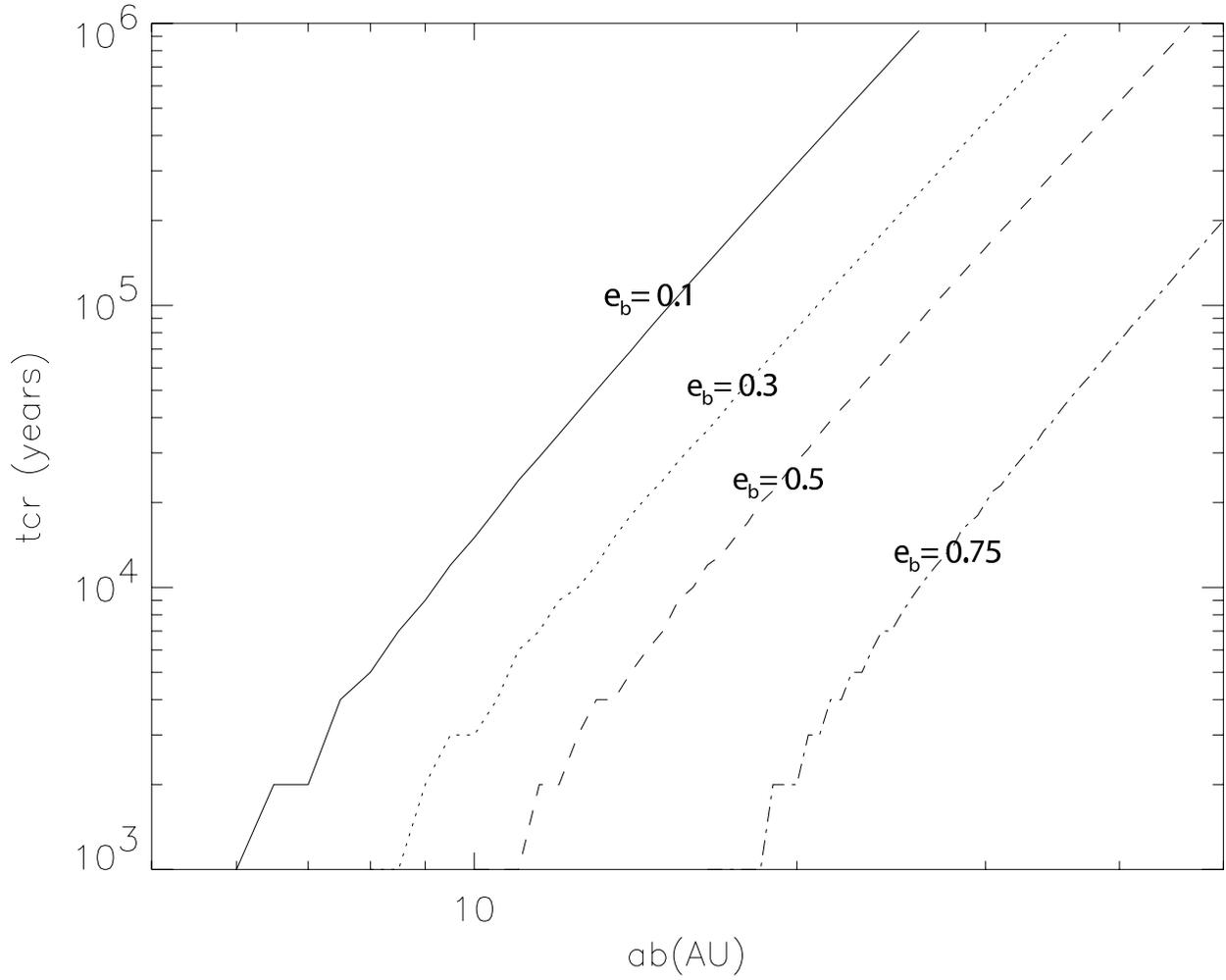}
\caption[]{Time for orbital crossing at 1 AU, as a function of the companion
semi--major axis, for 4 different values of $e_{b}$, and $m_{b}=0.5$} 
\label{tlab} 
\end{figure}

\begin{figure} 
\includegraphics[angle=0,origin=br,width=1.00\columnwidth]{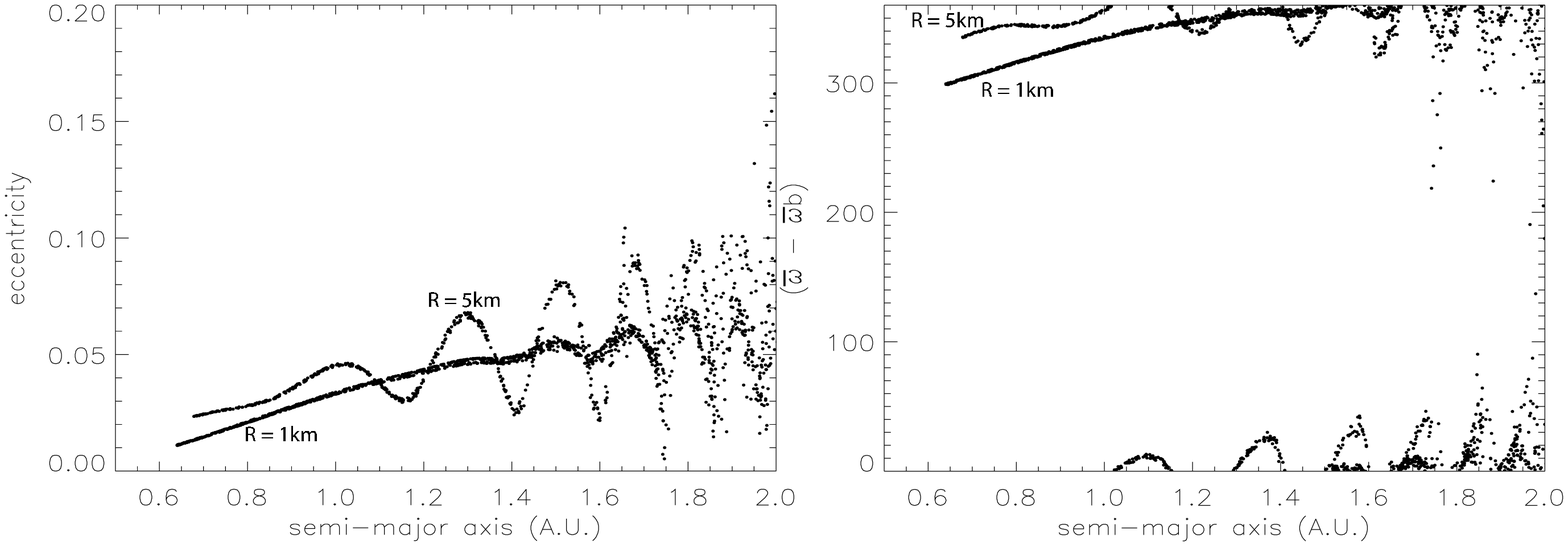}
\caption[]{Example gas drag run.
Snapshots, at $t=3\times10^{3}$yrs, of the ($e,a$) and ($\varpi-\varpi_b,a$)
distributions for 2 planetesimal populations of size $R_1=1\,$km and $R_2=5\,$km.
$\varpi-\varpi_b$ is the difference, in angular degrees,
between the particles and the companion star's longitude of periastron.
The companion star orbital parameters are: $a_b = 10\,$AU, $e_b = 0.3$, $m_b = 0.5$
 } 
\label{exsmall} 
\end{figure}

\begin{figure} 
\includegraphics[angle=0,origin=br,width=1.00\columnwidth]{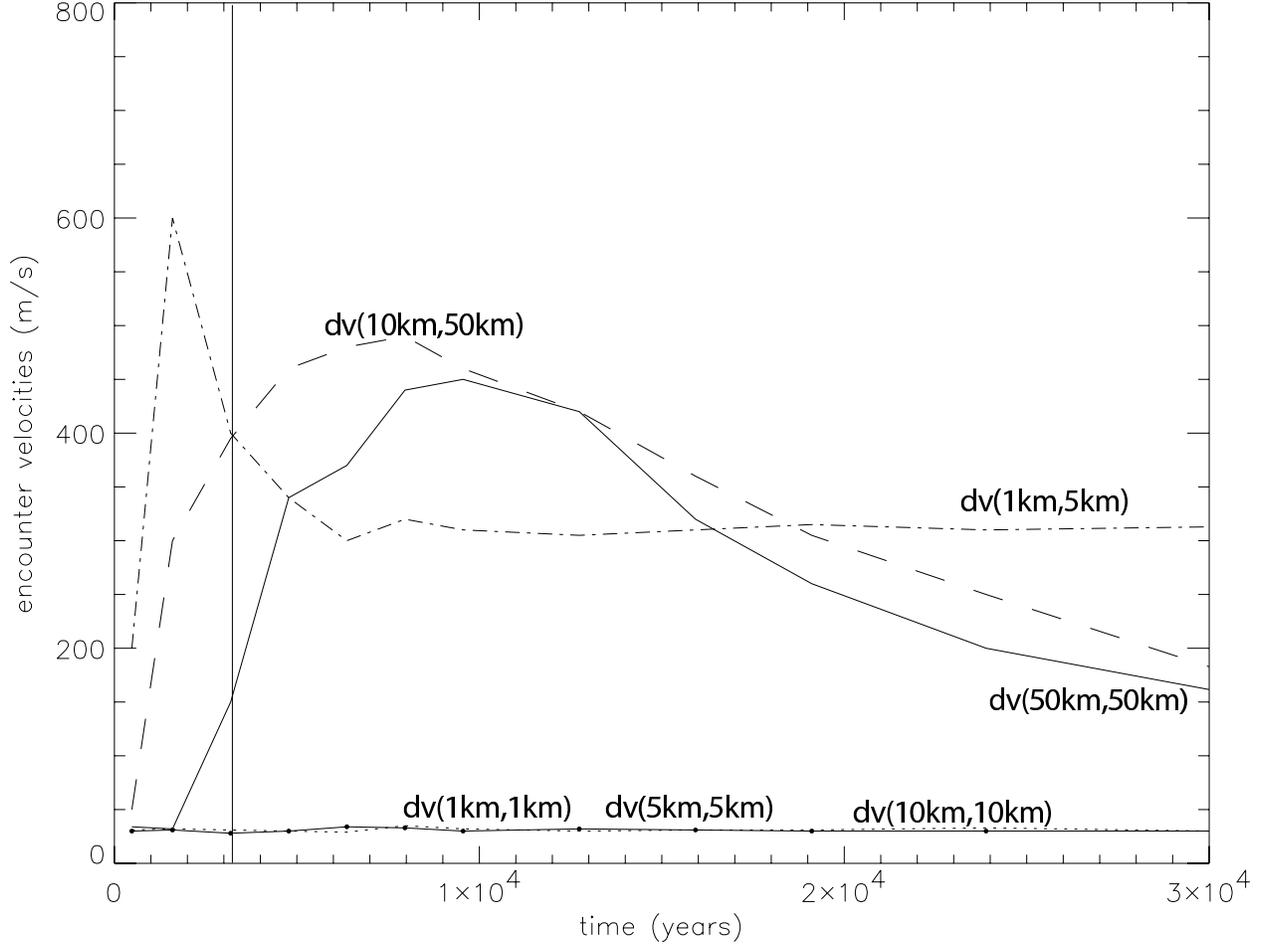}
\caption[]{Gas drag run.
Temporal evolution of mutual encounter velocities $\Delta v_{(R1,R2)}$
for different target-projectile pairs $R_1$ and $R_2$.
The vertical line shows the moment where orbital crossing should occur,
due to pure secular perturbations, according to Equ.\ref{tcr}.
Companion star orbital parameters: $a_b = 10\,$AU, $e_b = 0.3$, $m_b = 0.5$
 } 
\label{vitevol} 
\end{figure}

The situation is slightly more complex when considering a population of larger
objects. An important point is that gas drag is less efficient and no longer able
to systematically prevent orbital crossing, especially for the biggest
bodies (Fig.\ref{dv50-15}). 
As an example, for the 50\,km objects considered in Fig.\ref{dv50-15}, simulations
show that orbital crossing always occurs as predicted if $t_{cr}\leq 10^{4}$yrs
(the positions of the limit for orbital crossing approximately match that
of the $t_{cr}=10^{4}\,$yrs line in Fig.\ref{ebab})
but is prevented if it requires longer timescales (in this case, gas drag
has more time to affect the bodies orbital evolution).
It is worth noticing that
gas drag can never delay orbital crossing: either it prevents
it or it does not, and in this case crossing occurs at the time predicted
by Equ.\ref{ebab}. The only difference lies in the values of encounter velocities
after orbital crossing, which are always significantly lower than in the gas free
case. They remain however always high enough, i.e. $\geq 350$m.s$^{-1}$ for
50km bodies, to correspond to eroding impacts.
For $R_1 \neq R_2$ bodies, differential orbital phasing increases relative
velocities to values generally higher than for the small planetesimal case.
Nevertheless, the consequences of these high $\Delta v$ are less radical in
terms of accretion inhibition efficiency. If we divide the $(a_b,e_b)$
parameter space in a similar way as for the previous case, then we see
that the region of fully unperturbed encounters ($\Delta v\leq50$m.s$^{-1}$)
is slightly more extended. More interesting is maybe the extent of
the $50\leq\Delta v\leq250$m.s$^{-1}$ region, in which accretion
still prevails despite of the increased encounter velocities, and of the
``limbo'' zone  $250\leq\Delta v\leq1000$m.s$^{-1}$ with uncertain
accretion vs. erosion balance. Those 2 combined areas fill up most of the 
parameter space, leaving only a small fraction for without--doubt eroding
$\Delta v\geq1000$m.s$^{-1}$ impacts. Note also that these eroding
impacts almost always occur beyond the orbital crossing limit for
equal-sized 50\,km bodies.

\section{DISCUSSION}

\begin{figure} 
\includegraphics[angle=0,origin=br,width=1.00\columnwidth]{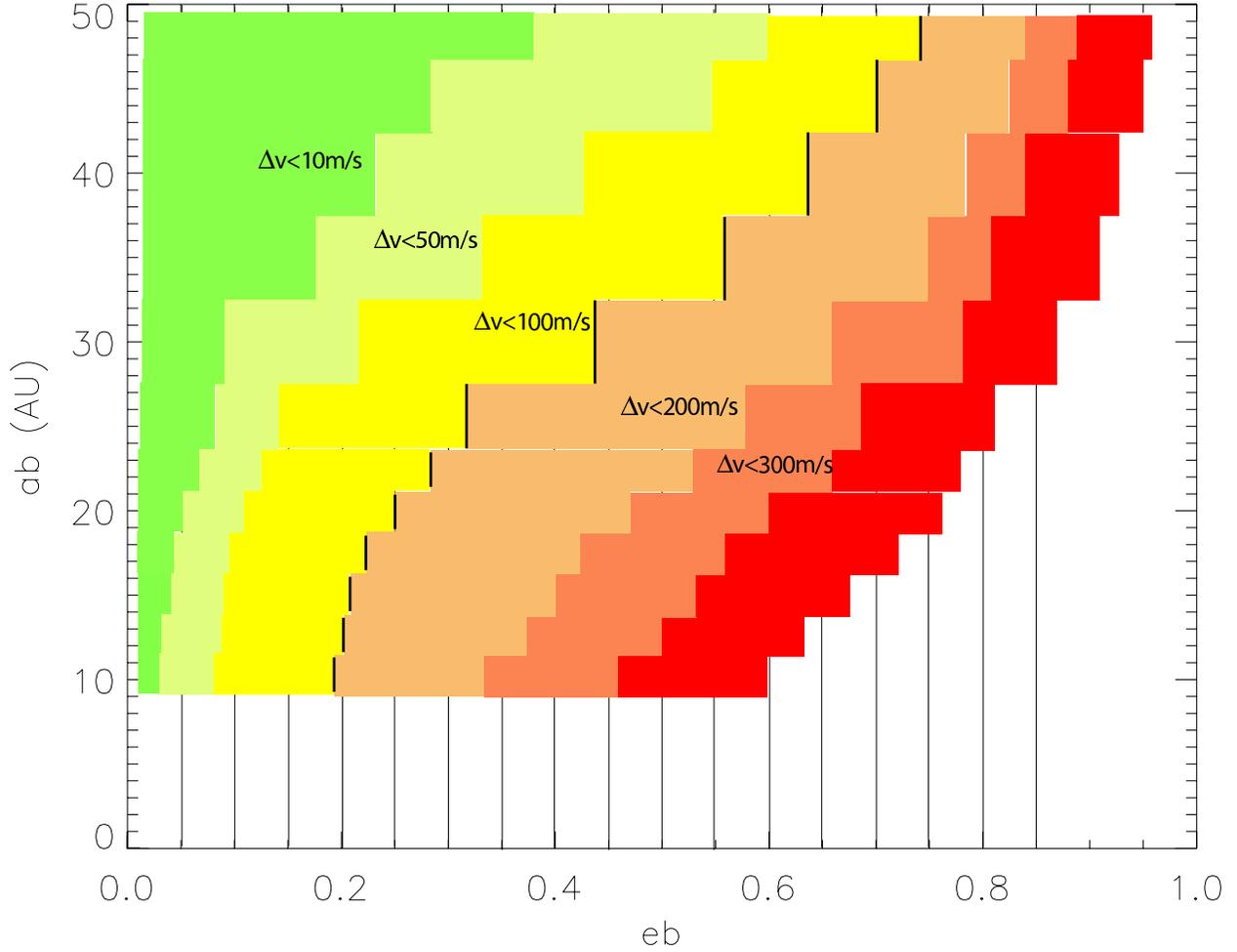}
\caption[]{Encounter velocities averaged, over the time interval
$0<t<2\times10^{4}$yrs, between $R_1=2.5$ and $R_2=5\,$km bodies
at 1 AU from the primary star, for different values of the
companion star semi-major axis and eccentricity.
The short black vertical segments mark the limit beyond which
$<\Delta v_{(R1,R2)}>$ values correspond to eroding impacts
for all tested collision outcome prescriptions.
 } 
\label{dv5-2.5} 
\end{figure}

\begin{figure} 
\includegraphics[angle=0,origin=br,width=1.00\columnwidth]{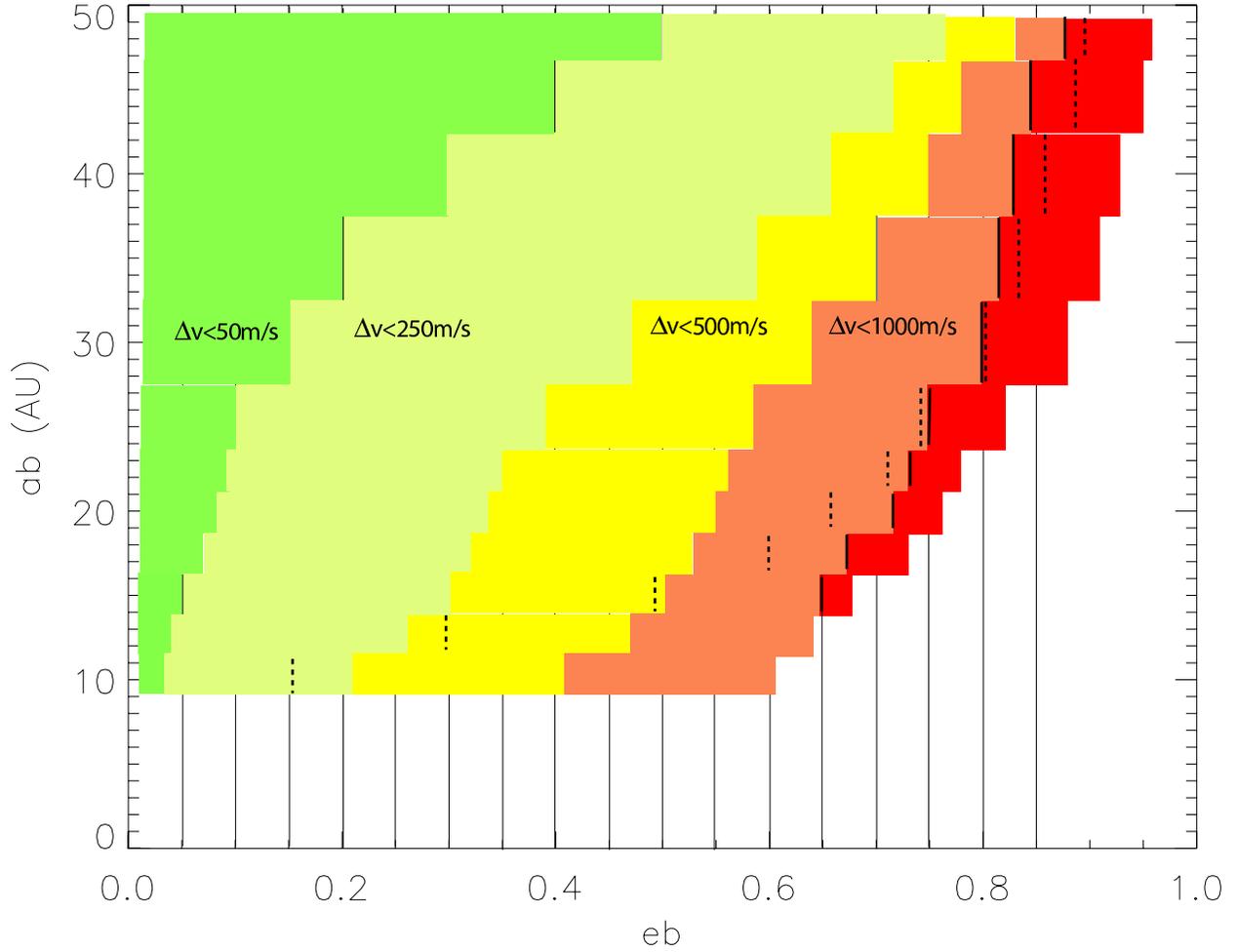}
\caption[]{Same as Fig.\ref{dv5-2.5}, but for $R_1=15$ and $R_2=50\,$km bodies.
The short black vertical segments mark the limit beyond which
$<\Delta v_{(R1,R2)}>$ values correspond to eroding impacts
for all tested collision outcome prescriptions.
The short dashed vertical segments mark the position beyond which orbital crossing
occurs for 50\,km bodies.
 } 
\label{dv50-15} 
\end{figure}

Comparing Figs.\ref{dv5-2.5} and \ref{dv50-15} to
Fig.\ref{ebab} for pure gravitational perturbing effects
clearly shows to what extent gas drag affects the encounter
velocity evolution.
On one hand, orbital alignment and eccentricity damping due
to gas friction tends to prevent orbital
crossing and even if it occurs, $\Delta v$ are lower than in the gas free case.
But on the other hand, gas drag introduces an additional
source of velocity increase
due to differential orbital alignment for bodies of different sizes.
This additional term proves to be very efficient, leading
to a $\Delta v$ increase for any departure from the exact
$R_1=R_2$ condition. Furthermore, this increase
is significant in a much larger region of the ($e_b$,$a_b$) phase space than
that delimited by the orbital-crossing limit in the gas free case
(Figs.\ref{dv5-2.5} and \ref{dv50-15}).
The global balance between these impacts for which
gas drag tends to slow down or prevent accretion, and
impacts where gas drag favours mutual growth,
i.e. those with 2 bodies of comparable sizes,
depends of course on the planetesimal initial size distribution,
and in particular the spread in planetesimal radius, a
parameter which is not very well constrained in current
planetary formation scenarios. This important issue is clearly beyond
the scope of the present study.
However, it is important to notice that large differential
phasing $\Delta v$ terms arise
for any small departure from the exact $R_1=R_2$ condition.
As an example, for most tested cases
a relative $\Delta R/R$ displacement of only 10\% from the equal-size configuration
results in a factor 2 (at least) increase in encounter velocities. As a consequence,
the low $\Delta v$ between equal--sized objects appears as a relatively
marginal result. It is likely that in a realistic planetesimal population,
with a statistical spread in object sizes, the dominant effect of gas
drag is the velocity increase due to differential orbital phasing.

To what extent does this velocity increase
affect the accretional evolution of the system? The answer to this
question depends mainly on the typical sizes within the ``initial'' planetesimal
swarm. Should this initial population be made of small ($\leq 10$\,km) objects,
then gas drag would probably present a major threat to accretion, at least
in the $a_b\leq 50\,$AU range explored here. 
Even taking our conservative criteria for fully eroding impacts,
Fig.\ref{dv5-2.5} clearly shows that the fraction of accretion--inhibiting
($e_b,a_b$) configurations is much higher than in the gas free case.
As an example, for pure secular perturbations most configurations
with $a_b\geq15\,$AU are accretion friendly when $e_b \la 0.7$,
whereas for gas drag and 5\,km planetesimals, in the same $a_b\geq15\,$AU
domain, the accretion/erosion frontier is approximately delimited by a
$(a_b-4{\rm AU})-60\,e_b\simeq 0\,\,$ line.
Furthermore, even if accretion is in principle possible in the ``limbo'' 
$10<\Delta v<100$m.s$^{-1}$ region of Fig.\ref{dv5-2.5}, it can not be of a
standard runaway type.
Indeed, runaway growth can only develop when encounter velocities
are significantly lower than the growing bodies escape velocity, increasing
their geometrical cross section by a
gravitational focusing factor proportional to $(v_{esc}/ \Delta v)^{2}$
\citep[e.g.][]{gre78}.
Such a focusing factor would be made negligible by the encounter velocity
increase in the ``limbo'' zone, where $\Delta v$ always exceed the escape
velocity of a $\leq10\,$km body.
This leaves only a very limited ($e_b,a_b$) region (the ``green'' area
of Fig.\ref{dv5-2.5}) where standard runaway accretion
can develop, i.e. where encounter velocities do not exceed the
escape velocities of typical $R<10\,$ km bodies.
For bigger $10<R<50\,$km objects, the situation is significantly different.
Here, differential orbital phasing almost never increases $\Delta v$
to values high enough to correspond to eroding impacts for all tested collision
prescriptions. For most cases, the limit for eroding impacts is actually given by
the limit for orbital crossing, thus probably making the latter mechanism the decisive
factor for accretion-inhibition. However, we are not back to a simple
black-or-white alternative as in a gas free case. Indeed, in most of the cases
below orbital crossing, $\Delta v_{R1 \neq R2}$ are increased to values exceeding the
escape velocity of a $50\,$km body. As already mentioned,
this would prevent the onset of 
standard runaway growth by decreasing the gravitational focusing factor. 
The ($a_b,e_b$) space for unperturbed runaway accretion is then restricted
to a region only marginally more extended than in the small planetesimals case.

In the extended ``limbo'' regions of both big and small planetesimals runs,
a possible accretion growth mode would be the ``type II'' runaway accretion
identified by \citet{kort01} in the context of giant planets' perturbations
on a swarm of planetesimals. Like in the present ``limbo'' zones, type II accretion
is characterized by initial encounter velocities increased by
differential gas--drag orbital phasing to values exceeding the biggest
objects' escape velocities, but without erosion overcoming accretion.
The first step of this scenario is
``orderly'' growth \citep{saf69} characterized by a slow and progressive growth
of all planetesimals. Growth later switches to runaway when the biggest bodies
reach a size large enough for gravitational focusing to become significant again.
For the specific case studied by \citet{kort01}, this size was approximately
that of Ceres. This turn--off size should of course vary with the
$m_b,a_b,e_b$ parameters of the perturber, but in a first approximation
Fig.\ref{dv50-15} shows that the escape
velocity of a Ceres--type object of radius $\simeq 500\,$km would indeed
exceed $\Delta v$ values in most of the parameter space ``left of''
the orbital crossing limit.

\section{LIMITATION TO THIS APPROACH AND PERSPECTIVES}

\subsection{Initial conditions}

One assumption made in our analytical derivations and
numerical simulations is to start from an initial system of objects
on circular orbits with very low relative velocities.
As already mentioned, this is implicitly equivalent to assuming that
the spread $\Delta t_0$
in the timescales for the formation of kilometre-sized planetesimal
is negligible compared to the characteristic timescales
for secular perturbations and planetesimal accretion. As discussed
in section 2, the validity of this assumption is directly linked to the
difficult problem of the mechanism for planetesimal formation.
This mechanism is still poorly understood and we believe our
initial conditions to be relatively reasonable within the frame of
the current dust-to-planetesimals formation process knowledge.

One has nevertheless to be aware of the possible effects
$\Delta t_0$ might have on the presented results. 
For the pure gravitational case, our assumption of initial
very small relative velocities breaks down if $\Delta t_0$ is
no longer negligible compared to $1/u$. In this case the first formed
planetesimals have the time to reach high eccentricities before
the later ones decouple from the gas, leading to an initial high
free $\Delta v$ component which cannot be damped
by later secular effects. Equ.\ref{realu} shows that the
typical limiting values for $\Delta t_0$ are in the
$\simeq 10^{3}$--$10^{4}\,$years range depending on the companion's orbital
parameter, in particular $a_b$.

\begin{figure} 
\includegraphics[angle=0,origin=br,width=1.00\columnwidth]{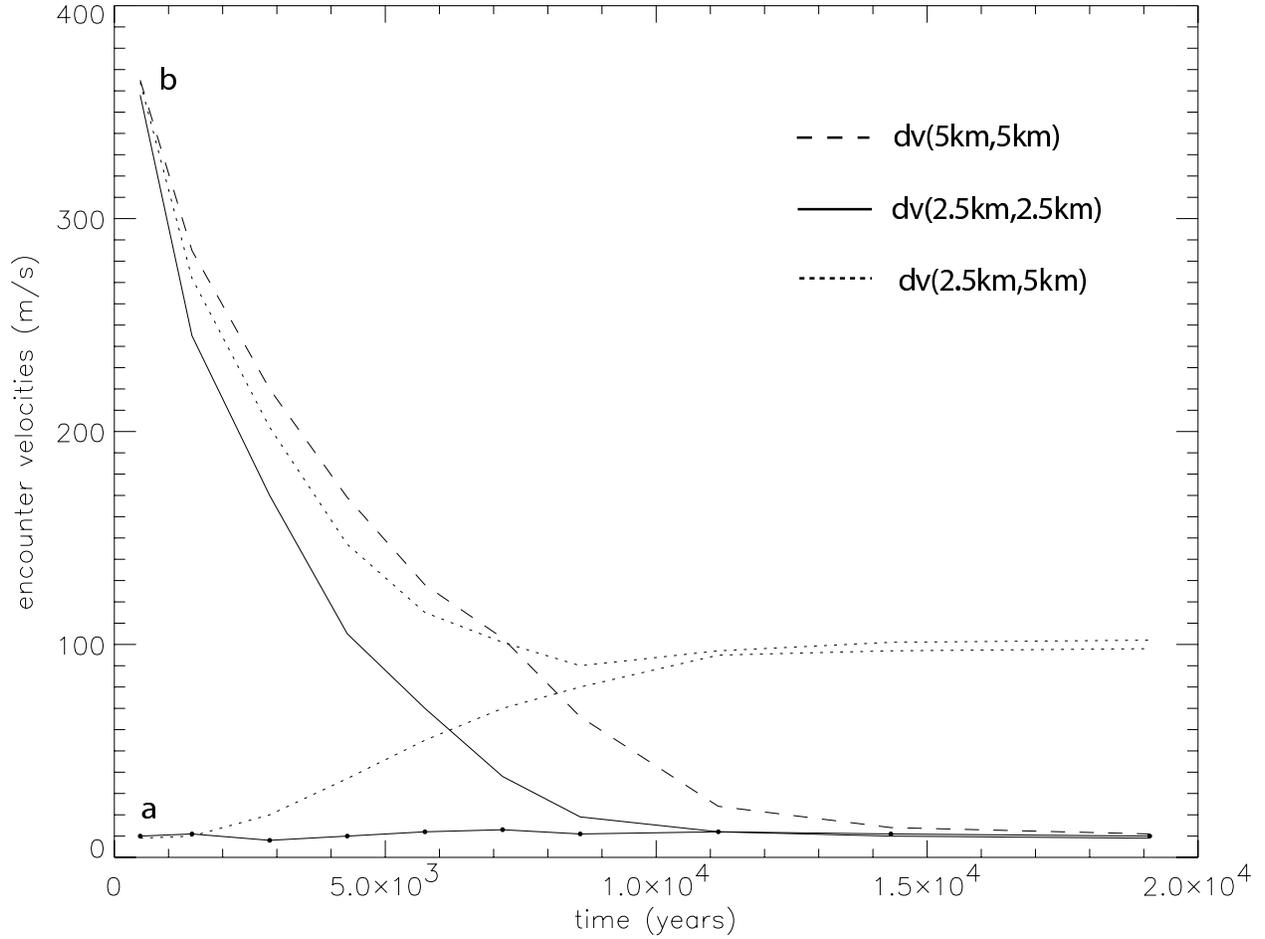}
\caption[]{Average encounter velocities, at 1 AU from the primary, between
planetesimals of different sizes for 2 different initial conditions:
a) circular orbits
b) initial eccentricities randomly distributed between 0 and
the equilibrium phased value.
The companion's orbital parameter are $a_b = 30\,$AU, $e_b = 0.4$, $m_b = 0.5$
 } 
\label{vitconv} 
\end{figure}

In the gas drag runs, the effect of an initial $\Delta t_0$ is
less critical, since gas friction tends to progressively
damp any initial orbital differences toward the same equilibrium
value. This is clearly illustrated in the example displayed
in Fig.\ref{vitconv}, where we artificially introduce an initial $\Delta t_0$
by allowing initial eccentricities to vary between 0 and
the value they would reach at the end of the gas drag induced phasing
(thus implicitly assuming that some objects appeared earlier than others
and had enough time to be affected by the coupled effect of secular perturbations and
gas drag). As can be seen, the initial high $\Delta v_{R1,R2}$
values are progressively damped and after a transition period of less than
$\simeq 10^{4}\,$years converge towards the same equilibrium values
than those obtained when starting from circular orbits.
The duration of this critical transition period is of course
dependent of the companion star orbital parameters and of planetesimal sizes,
and it might in some cases be long enough to significantly affect
the accretion process.
Possible effects of initial $\Delta t_0$ can thus not be ruled out
within the frame of our present knowledge of the planetesimal
formation process.
It is however important to note that any initial $\Delta t_0$ would
necessarily act 
in the direction of accretion $inhibition$ by adding an additional
$\Delta v$ term. Our results might thus be considered as a lower limit for 
perturbing effects within binaries.

\subsection{Gas friction modeling}

Our gas drag model is a simplified one where the gas disk
is assumed to be fully axisymmetric and follows a classical
\citet{haya81} power law distribution. It is however more
than likely that in reality the gas disk should depart
from this simplified view because it would also "feel" the
companion star's perturbations. Several numerical studies
have investigated the complex behaviour of gaseous disks in binary
systems. They all show that pronounced spiral structures
rapidly form within the disk \citep[e.g.][]{arty94,savo94}
and that gas streamlines exhibit radial velocities.
To follow the dynamical behaviour of planetesimals in such non-axisymmetric
gas profiles would require a study of the $coupled$ evolution
of both gas and planetesimal populations, which 
would probably have to rely on hydro--code modeling
of the gas in addition to N--body type models for the planetesimals.
Such an all--encompassing gas$+$planetesimals modeling is clearly the next step
in binary disk studies.

\subsection{Do planetesimals really form?}

One implicit assumption behind our simulations is of course that,
as some point, kilometre-sized planetesimals $do$ form in the system.
With our present understanding of the planet--formation--in--binaries
problem, nothing allows us to be sure that this assumption is valid. 
The crucial problem of how dust sticking and/or dust settling and
gravitational instabilities can proceed in a binary system remains yet
to be investigated.
Problems could also arise at even earlier stages. \citet{nels00}
has for instance shown that, under certain conditions, thermal energy
dissipation in gas disks for equal mass binaries could inhibit
dust coagulation by raising temperatures above vaporization limit.
But, as recognized by the author himself,
these results are still preliminary, and a complete study of this problem,
taking into account a broader range of physical parameters, has yet
to be carried out.

\section{SUMMARY}

We explore the effect of gravitational perturbations of a
stellar companion of semi--major axis $a_b\leq50\,$AU
on a planetesimal population orbiting the primary star,
with particular focus on the 1\,AU region.
We concentrate on the relative velocity distribution, 
the key parameter for the evolution of a planetesimal
population, which can only accrete each other if $\Delta v$
remain below a threshold value which depends on planetesimal sizes.
We investigate the companion's influence on this
distribution for timescales comparable to that of the standard
runaway growth scenario.

We first address the simplified pure gravitational problem.
In this case, the companion stars triggers strong phased secular orbital
oscillations which might eventually become so narrow
that neighboring orbits cross, leading to abrupt encounter
velocity increase. We derive a set of semi-analytical expressions
to determine this orbital crossing location as a function
of time. Assuming initial
coapsidality for planetesimal orbits at 1AU from the primary,
we find that for a large fraction
of companion orbital parameter configurations, schematically
for $a_b\geq 15\,$AU and $e_b\leq 0.6$--$0.7$,
orbital crossing does not occur within typical runaway
timescales of $\simeq 10^{4}\,$years. For the cases where orbital
crossing occurs, however, impacts velocities reach values almost
always too high to allow mutual accretion.

The inclusion of gas drag greatly complicates this 
black-or-white picture and leads to a large 
spectrum of impact velocities between planetesimals depending on 
their size and on the binary orbital parameters. In our numerical
exploration we observe the well known effect of forced
orbital alignment between equal-sized objects
which prevents orbital crossing from occurring for most planetesimal
sizes explored here. But this effect is balanced by the
$\Delta v$ increase between objects of {\rm different sizes} due to
the strong size dependency of the gas drag induced orbital phasing.
In fact, we find that this differential phasing term is so sensitive to any small
departure from the exact equal--size case that it is
likely to be the dominant effect in any "real" planetesimal population
with even a limited  spread in planetesimal sizes.
For a typical companion mass $m_b=0.5$ and for the
range of semi-major axis here explored ($a_b\leq50\,$AU) we find
that "standard" runaway accretion, where $\Delta v$ remain smaller
than bodies' typical escape velocities $v_{esc}$, is only possible for companion
eccentricities close to 0.
In the rest of the ($e_b$,$a_b$) parameter space, impact velocities always
exceed $v_{esc}$ and might even exceed the threshold values $v_{ero}$ for which
the net balance for collision outcomes is erosion instead of accretion.
We can thus basically divide the ($e_b$,$a_b$) parameter space in 3 regions
\begin{itemize} 
\item The domain where  $\Delta v \leq v_{esc}$. Here we expect the companion's
perturbation to have a negligible influence on the accretion evolution of
the system. This region is relatively narrow, for all planetesimal sizes explored,
for $a_b\leq40\,$AU.
\item The region where $\Delta v$ exceeds the erosion threshold value
for any realistic collision outcome prescriptions we tested.
The extent of this region depends on the planetesimal sizes. It is
relatively wide for ``small'' $\leq 10\,$km bodies, where this
accretion inhibition mechanism is much more efficient than orbital
crossing in the gas free case. For objects in the $\simeq 50\,$km size range,
the $\Delta v \geq v_{ero}$ condition is met in a more limited ($e_b$,$a_b$)
domain. Besides, it occurs mostly for configurations where secular orbital crossing
occurs despite gas friction, so that the latter mechanism is probably
the dominant one for inhibiting accretion of large objects.
\item In between these 2 well defined domains, there is an intermediate
region where accretion is possible despite $\Delta v$ values exceeding $v_{esc}$,
or where the erosion vs. accretion net balance is uncertain, 
i.e. it depends on the assumed collisional outcome model.
Here, a possible accretion evolution scenario could be the type II runaway growth
identified by \citet{kort01}, where accretion starts in a slow orderly way,
and only later switches to runaway when larger planetesimals have formed.
\end{itemize}
 
.

{\bf Acknowledgements}

The Authors wish to thank Eiichiro Kokubo and an anonymous referee
for fruitful comments and suggestions that helped improve the paper a lot

\clearpage

{} 

\clearpage
\begin{table} 
\caption[]{Average encounter velocities in m.s$^{-1}$, at 1\,AU from the primary,
within a population of ``small'' planetesimals $1<R<10\,$km for a
gas drag simulation with the companion star parameters
$m_b=0.5$, $a_b=10\,$AU and $e_b=0.3$. $\Delta v_{R1,R2}$ values are
averaged over the time interval $0<t<2\times10^{4}$yrs.
Initial starting encounter velocities are such as $\Delta v_0\simeq 10$m.s.$^{-1}$.
$\Delta v$ values in bold correspond to accreting impacts for
all tested collision outcome prescriptions. Underlined values
are those for which we obtain different accretion vs. erosion
balance depending on the tested prescription. Values in classical roman
characters correspond to cases for which all tested models
agree on a net erosive outcome.
}
\label{dvmatrix1a} 
\begin{tabular}{c | cccccccccc}\\
\hline 
Sizes (km) & 1 & 2 & 3 & 4 & 5 & 6 & 7 & 8 & 9 & 10\\ 
\hline 
1 & {\bf10} & 154 & 233 & 285 & 327 & 360 & 391 & 426 & 452 & 458\\ 
2 & 172 & {\bf10} & 94  & 133 & 187 & 223 & 262 & 287 & 316 & 334\\ 
3 & 238 & 84 & {\bf11}  & 54  & 99 & 137 & 177 & 200 & 230 & 254\\ 
4 & 289 & 144 & 63  & {\bf12} &\underline{40}& 80  & 115 & 149 & 171 & 198\\ 
5 & 325 & 188 & 103 &\underline{43}& {\bf12} & \underline{32}& \underline{70}& 100 & 122 & 154\\ 
6 & 373 & 228 & 144 & 83  & \underline{32}& {\bf11} & {\bf36}& \underline{56}& 84  & 104\\ 
7 & 400 & 261 & 182 & 113 & \underline{68}& {\bf36}& {\bf12} & {\bf35}  &\underline{48} & \underline{76}\\ 
8 & 428 & 298 & 212 & 147 & 98 & \underline{56}&{\bf35}& {\bf12} &{\bf36}& \underline{45}\\ 
9 & 450 & 310 & 238 & 168 & 123 & 83  & \underline{48}& {\bf36}& {\bf13} & {\bf31}\\ 
10& 453 & 338 & 263 & 196 & 152 & 107 & \underline{73}& \underline{48}& {\bf31} & {\bf13}\\ 
\end{tabular} 
\end{table}

\clearpage
\begin{table} 
\caption[]{Same as Table \ref{dvmatrix1a}, but for a population of bigger
planetesimals $10<R<50\,$km.
Initial starting encounter velocities are such as $\Delta v_0\simeq 30$m.s.$^{-1}$
}
\label{dvmatrix1b} 
\begin{tabular}{c | ccccccccc}\\
\hline 
Size (km)&10& 15     & 20     & 25     & 30     & 35     & 40     & 45     & 50\\ 
\hline 
10 &{\bf28} &\underline{100} &\underline{163}&\underline{216}&\underline{261}&\underline{296}&\underline{329}&\underline{364}&{\bf385}\\ 
15 &\underline{99} &{\bf29} &\underline{109} &\underline{149}&\underline{204}&\underline{243}&\underline{277}&\underline{314}&{\bf345}\\ 
20 &\underline{170}&\underline{108} &\underline{133}&\underline{128} &\underline{177} &\underline{223}&\underline{249}&\underline{289}&\underline{324}\\ 
25 &\underline{215}&\underline{155}&\underline{140} &\underline{158} &\underline{183} &\underline{211} &\underline{224}&\underline{274}&\underline{304}\\ 
30 &\underline{263}&\underline{208}&\underline{185} &\underline{183} &\underline{221} &\underline{238} &\underline{247} &\underline{283} &\underline{316}\\ 
35 &\underline{299}&\underline{250}&\underline{218}&\underline{208} &\underline{237} &\underline{251} &\underline{253} &\underline{273} &\underline{311}\\ 
40 &\underline{337}&\underline{282}&\underline{259}&\underline{225} &\underline{257} &\underline{274} &\underline{257} &\underline{299} &\underline{335}\\ 
45 &\underline{365}&\underline{320}&\underline{291}&\underline{261}&\underline{277} &\underline{279} &\underline{276} &\underline{341} &\underline{338}\\ 
50 &{\bf387}&{\bf339}&\underline{321}&\underline{299}&\underline{303}&\underline{293} &\underline{314} &\underline{332} &\underline{356}\\ 
\end{tabular} 
\end{table}

\clearpage
\begin{table} 
\caption[]{Same as Table \ref{dvmatrix1a}, i.e. ``small'' planetesimal
population, but with companion
star parameters $m_b=0.5$, $a_b=20\,$AU and $e_b=0.4$.
}
\label{dvmatrix2a} 
\begin{tabular}{c | cccccccccc}\\
\hline 
Sizes (km) & 1 & 2 & 3 & 4 & 5 & 6 & 7 & 8 & 9 & 10\\ 
\hline 
1 & {\bf11} & 127 & 204 & 255 & 298 & 342 & 368 & 390 & 417 & 442\\ 
2 & 126 & {\bf10} & 84  & 139 & 185 & 227 & 258 & 290 & 317 & 340\\ 
3 & 200 & 91 & {\bf11}  & 68  & 108 & 158 & 186 & 218 & 246 & 272\\ 
4 & 258 & 146 & 61  & {\bf9} &\underline{48}& 88  & 120 & 154 & 186 & 209\\ 
5 & 301 & 192 & 113 & 54  & {\bf12} & \underline{44}& \underline{75}& 111 & 136 & 164\\ 
6 & 339 & 232 & 152 & 99  & \underline{43}& {\bf10} & {\bf31}& \underline{66}& 92  & 119\\ 
7 & 361 & 262 & 181 & 126 & \underline{77}& {\bf28}& {\bf13} & {\bf26}  &\underline{56} & 87\\ 
8 & 395 & 295 & 219 & 159 & 112 & \underline{68}&{\bf26}& {\bf11} &{\bf28}& \underline{48}\\ 
9 & 425 & 320 & 246 & 190 & 136 & 92  & \underline{55}& {\bf25}& {\bf11} & {\bf23}\\ 
10& 446 & 346 & 266 & 208 & 163 & 122 & 82  & \underline{49}& {\bf22} & {\bf12}\\ 
\end{tabular} 
\end{table}

\clearpage
\begin{table} 
\caption[]{Same as Table \ref{dvmatrix2a} but for the ``big'' planetesimal population.
}
\label{dvmatrix2b} 
\begin{tabular}{c | ccccccccc}\\
\hline 
Size (km)&10& 15     & 20     & 25     & 30     & 35     & 40     & 45     & 50\\ 
\hline 
10 &{\bf26} &\underline{95} &\underline{154}&\underline{198}&\underline{231}&\underline{265}&\underline{292}&\underline{312}&{\bf335}\\ 
15 &\underline{92} &{\bf27} &{\bf73} &\underline{115}&\underline{148}&\underline{179}&\underline{210}&\underline{234}&{\bf268}\\ 
20 &\underline{157}&{\bf75} &{\bf32} &{\bf51} &{\bf88} &\underline{125}&\underline{146}&{\bf170}&{\bf208}\\ 
25 &\underline{197}&\underline{111}&{\bf54} &{\bf29} &{\bf48} &{\bf74} &{\bf101}&{\bf121}&{\bf145}\\ 
30 &\underline{235}&\underline{149}&{\bf90} &{\bf47} &{\bf28} &{\bf40} &{\bf63} &{\bf87} &{\bf101}\\ 
35 &\underline{266}&\underline{184}&\underline{117}&{\bf75} &{\bf42} &{\bf30} &{\bf37} &{\bf56} &{\bf73}\\ 
40 &\underline{290}&\underline{208}&\underline{145}&{\bf97} &{\bf65} &{\bf40} &{\bf32} &{\bf36} &{\bf53}\\ 
45 &\underline{308}&\underline{226}&{\bf167}&{\bf120}&{\bf86} &{\bf59} &{\bf37} &{\bf31} &{\bf33}\\ 
50 &{\bf341}&{\bf272}&{\bf207}&\bf{144}&{\bf104}&{\bf77} &{\bf56} &{\bf38} &{\bf35}\\ 
\end{tabular} 
\end{table}
%


\clearpage

\clearpage

\clearpage

%


\clearpage

%


\clearpage

%


\clearpage

%

\clearpage

%
\clearpage

%


\clearpage

%

\clearpage

\end{document}